\documentclass{aa}  
\usepackage{graphicx}
\usepackage{txfonts}
\usepackage[colorlinks=true,citecolor=blue]{hyperref}
\usepackage{multirow}
\usepackage{amsmath,amstext}
\usepackage[T1]{fontenc}
\usepackage{color}
\usepackage{comment}
\usepackage{caption}
\DeclareCaptionFormat{cont}{#1 (cont.)#2#3\par}
\usepackage[flushleft]{threeparttable}

\DeclareRobustCommand{\ion}[2]{%
\relax\ifmmode
\ifx\testbx\f@series
{\mathbf{#1\,\mathsc{#2}}}\else
{\mathrm{#1\,\mathsc{#2}}}\fi
\else\textup{#1\,{\mdseries\textsc{#2}}}%
\fi}

\newcommand{\orcid}[1]{\href{https://orcid.org/#1}{\includegraphics[width=10pt]{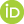}}}
\newcommand{\github}[1]{\href{https://github.com/#1}{\includegraphics[width=10pt]{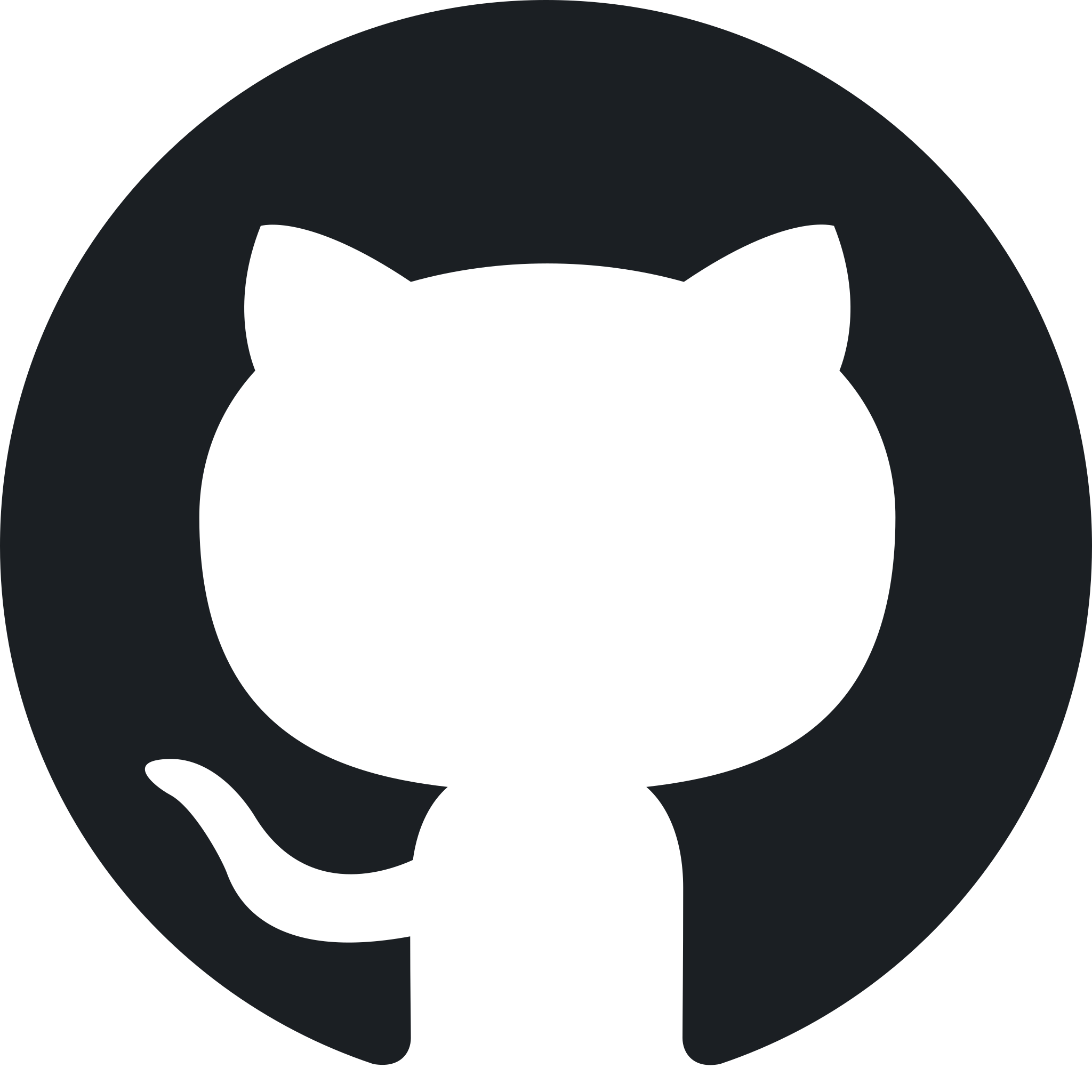}}}

\newcommand{\cspi}{CSP-I\xspace}
\newcommand{\cspii}{CSP-II\xspace}
\newcommand{\snoopy}{SNooPy\xspace}
\newcommand{\pisco}{PISCOLA\xspace}
\newcommand{\bayesn}{BayeSN\xspace}

\newcommand{\maxmodel}{\textit{max\_model}\xspace}
\newcommand{\EBVmodel}{\textit{EBV\_model2}\xspace}

\newcommand{\tmax}{$T_{\rm max}$\xspace}
\newcommand{\sbv}{$s_{BV}$\xspace}
\newcommand{\dtmax}{$\Delta$\tmax}
\newcommand{\gmax}{$g_{\rm max}$\xspace}
\newcommand{\rmax}{$r_{\rm max}$\xspace}
\newcommand{\Jmax}{$J_{\rm max}$\xspace}
\newcommand{\Hmax}{$H_{\rm max}$\xspace}
\newcommand{\J}{\textit{J}\xspace}
\renewcommand{\H}{\textit{H}\xspace} 

\newcommand{\hunits}{km\,s$^{-1}$\,Mpc$^{-1}$\xspace}

\begin{document} 
\title{Testing the Homogeneity of Type Ia Supernovae in the Near-Infrared for Accurate Distance Estimations}
\titlerunning{Testing the homogeneity of SNe~Ia in the NIR}
\author{
T.~E. M\"uller-Bravo\inst{1}\thanks{\email{t.e.muller-bravo@ice.csic.es}}\orcid{0000-0003-3939-7167},
L. Galbany\inst{1,2}\orcid{0000-0002-1296-6887},
E. Karamehmetoglu\inst{3}\orcid{0000-0001-6209-838X},
M. Stritzinger\inst{3}\orcid{0000-0002-5571-1833},
C. Burns\inst{4}\orcid{0000-0003-4625-6629},\newline
K. Phan\inst{3,2}\orcid{0000-0001-6383-860X},
A. I\'a\~nez Ferres\inst{1}, 
J.~P. Anderson\inst{5}\orcid{0000-0003-0227-3451},
C. Ashall\inst{6}\orcid{0000-0002-5221-7557},
E. Baron\inst{7,8,9}\orcid{0000-0001-5393-1608},
P.~Hoeflich\inst{10}\orcid{0000-0002-4338-6586},
E.~Y. Hsiao\inst{10}\orcid{0000-0003-1039-2928 },\newline
T. de~Jaeger\inst{6}\orcid{0000-0001-6069-1139},
S. Kumar\inst{10},
J. Lu\inst{10}\orcid{0000-0002-3900-1452},
M.~M. Phillips\inst{11}\orcid{0000-0003-2734-0796},
M. Shahbandeh\inst{10}\orcid{0000-0002-9301-5302},\newline
N. Suntzeff\inst{12}\orcid{0000-0002-8102-181X} and
S.~A. Uddin\inst{12,13}\orcid{0000-0002-9413-4186}
}
\authorrunning{M\"uller-Bravo et al.}
\institute{
Institute of Space Sciences (ICE, CSIC), Campus UAB, Carrer de Can Magrans, s/n, E-08193 Barcelona, Spain
\and
Institut d’Estudis Espacials de Catalunya (IEEC), E-08034 Barcelona, Spain
\and
Department of Physics and Astronomy, Aarhus University, Ny Munkegade 120, DK-8000 Aarhus C, Denmark
\and
Observatories of the Carnegie Institution for Science, 813 Santa Barbara St., Pasadena, CA 91101, USA
\and
European Southern Observatory, Alonso de C\'ordova 3107, Casilla 19, Santiago, Chile
\and
Institute for Astronomy, University of Hawai'i, 2680 Woodlawn Drive, Honolulu, HI 96822, USA
\and
Department of Physics \& Astronomy, University of Oklahoma, Norman, OK 73019 USA
\and
Department of Physics \& Astronomy, George Washington University, Washington, DC USA
\and
Hamburger Sternwarte, Hamburg, Germany
\and
Department of Physics, Florida State University, 77 Chieftan Way, Tallahassee, FL 32306, USA
\and
Carnegie Observatories, Las Campanas Observatory, Casilla 601, La Serena, Chile
\and
George P. and Cynthia Woods Mitchell Institute for Fundamental Physics and Astronomy, Texas A\&M University, Department of Physics and Astronomy,  College Station, TX 77843, USA
\and 
Centre for Space Studies, American Public University System, 111 W. Congress Street, Charles Town, WV 25414, USA
}
\date{Received \today; accepted ---}

\abstract
{Since the discovery of the accelerating expansion of the Universe more than two decades ago, Type Ia Supernovae (SNe~Ia) have been extensively used as standardisable candles in the optical. However, SNe~Ia have shown to be more homogeneous in the near-infrared (NIR), where the effect of dust extinction is also attenuated. In this work, we explore the possibility of using a low number of NIR observations for accurate distance estimations, given the homogeneity at these wavelengths. We found that one epoch in \J and/or \H band, plus good \textit{gr}-band coverage, gives an accurate estimation of peak magnitudes in \J (\Jmax) and \H (\Hmax) bands. The use of a single NIR epoch only introduces an additional scatter of $\sim0.05$\,mag for epochs around the time of \textit{B}-band peak magnitude (\tmax). We also tested the effect of optical cadence and signal-to-noise ratio (S/N) in the estimation of \tmax and its uncertainty propagation to the NIR peak magnitudes. Both cadence and S/N have a similar contribution, where we constrained the introduced scatter of each to $<0.02$\,mag in \Jmax and $<0.01$ in \Hmax. However, these effects are expected to be negligible, provided the data quality is comparable to that obtained for observations of nearby SNe ($z \lesssim 0.1$). The effect of S/N in the NIR was tested as well. For SNe~Ia at $0.08<z\lesssim0.1$, NIR observations with better S/N than that found in the CSP sample is necessary to constrain the introduced scatter to a minimum ($\lesssim0.05$\,mag). These results provide confidence for our FLOWS project that aims in using SNe~Ia with public ZTF optical light curves and few NIR epochs to map out the peculiar velocity field of the local Universe. This will allow us to determine the distribution of dark matter in our own supercluster, Laniakea, and test the standard cosmological model by measuring the growth rate of structures, parameterised by $fD$, and the Hubble-Lemaître constant, $H_0$. All of the software developed and used throughout this work is publicly available.\github{HOSTFLOWS/flows_sims}}
\maketitle

\section{Introduction}

The expansion rate of the Universe, parameterised by the Hubble-Lemaître parameter, H$(z)$,  varies across cosmic time. In the last few years, there has been tremendous effort to measure the local value, known as the Hubble-Lemaître constant ($H_0$), with extremely high precision \citep[$<2\%$ uncertainty;][]{Riess2021}. Recent results have further increased the discrepancy in the value of $H_0$ between the local distance ladder \citep[$H_0 =$ 73.04 $\pm$ 1.04 \hunits, baseline with systematics;][]{Riess2021} and the cosmic microwave background \citep[CMB; $H_0 =$ 67.4 $\pm$ 0.5 \hunits;][]{Planck2020} measurements, colloquially known as the ``Hubble tension\rq\rq, to 5$\sigma$ (although see \citealt{Freedman2019, Huang2020, Khetan2021} for some alternative local measurements). This discrepancy possibly hints towards new physics beyond the standard cosmological model, or alternatively, unaccounted systematic effects (see \citealt{DiValentino2021} for a recent review on the Hubble tension).

In the local Universe, the recession velocities measured from galaxies are affected by a combination of the expansion of the Universe and the gravitational pull of other adjacent galaxies. The measurement of these peculiar velocities is critical for two main reason. First, cosmological analyses with Type~Ia Supernovae (SNe~Ia) rely on discerning the contribution of peculiar velocities to isolate the cosmological redshift. Secondly, peculiar velocities can be used to inferred the matter-density distribution in the local Universe \citep{Peebles1976}, including our own supercluster, Laniakea \citep[e.g.,][]{Tully2014}. The latter provides a direct measurement of the growth-rate of structure, which can be compared to estimates from the early Universe \citep[e.g.,][]{Linder2005}.

Current measurements of peculiar velocities often rely on methods such as the Fundamental-Plane and Tully-Fisher relations \citep[e.g.,][]{Tully2016}, which provide distances with relatively large uncertainties (root-mean-squared [RMS] of $\sim20-30$\% per galaxy) and only reach out to $z\sim0.05$, preventing the study of the peculiar velocities at further distances. Therefore, there is a lack of higher-precision methods that can also extend to further redshifts for the estimation of distances in the local Universe.

Since the discovery of the accelerating expansion of the Universe more than two decades ago \citep{Riess1998, Perlmutter1999}, SNe~Ia have been extensively used as cosmological distance indicators. In the optical, their light curves can be standardised through empirical relations between their peak brightness, stretch \citep[e.g.,][]{Rust1974, Pskovskii1977, Phillips1993} and colour \citep{Tripp1998}. In addition, SNe~Ia are brighter in the optical (where detectors are larger as well), compared to other wavelengths, making them easier to observe. Therefore, cosmological analyses with SNe~Ia \citep[e.g.,][]{Betoule2014, Scolnic2018, Abbott2019} commonly focus on optical wavelengths and rely on light-curve fitters, such as SALT2 \citep{Guy2005, Guy2007}, for the estimation of their light-curve parameters. Moreover, SNe~Ia in the optical have recently been used to estimate the growth-rate of structures \citep[e.g.,][]{Boruah2020, Stahl2021}.

SNe~Ia were first proposed as distance indicators in the near-infrared (NIR) around four decades ago \citep[][but also see \citealt{Meikle2000}]{Elias1981, Elias1985}, where they seem to be true \textit{standard candles} (as opposed to \textit{standardisable candles} in the optical). In other words, an estimation of the NIR peak magnitudes is all that is needed to measure distances. The NIR light curves of SNe~Ia present lower intrinsic dispersion than the optical light curves and have the advantage of being less affected by dust extinction, which makes them exceptional for measuring cosmological distances \citep[e.g.,][]{Krisciunas2004, Wood-Vasey2008, Freedman2009, Barone-Nugent2012, Phillips2012, Weyant2014, Friedman2015}. Moreover, the NIR light curves of SNe~Ia have already been used to constrain $H_0$ to a few percent \citep[e.g.,][]{Burns2018, Dhawan2018}.

The low intrinsic dispersion of SNe~Ia in the NIR raises the the possibility of using them to achieve accurate cosmography by measuring peculiar velocities of local galaxies, reaching out to $z \sim 0.1$ or even beyond. However, the sample of SNe~Ia observed in the NIR is currently low due to several factors: low NIR detector sensitivity in the past; SNe~Ia are fainter at these wavelengths, where the sky brightness dramatically decreases the contrast for the (SN) observations, thus needing to integrate for longer; and the number of facilities with NIR instruments (compared to optical ones) is low. However, by taking advantage of the exceptional homogeneity of the SNe~Ia in the NIR, we can possibly reconstruct their light curves with just a few photometric data points \citep[e.g.,][]{Krisciunas2004}, increasing the total number of observed objects.

Given the large stream of optical photometry publicly provided by the Zwicky Transient Facility \citep[ZTF;][]{Graham2019}, which works as a precursor and testing ground for LSST, hundreds to thousands of SNe~Ia are being followed-up with high-cadence (average of 2 days) \textit{gr}-bands photometry\footnote{\url{https://www.ztf.caltech.edu/ztf-public-releases.html}}. Thus, ZTF can provide the optical data coverage while NIR photometry can be obtained with other facilities. This work aims to test how accurately we can retrieve NIR peak magnitudes with well-covered optical light curves and few NIR epochs for distance estimations. Our results will give assurance to use SNe~Ia with public ZTF \textit{gr}-band light curves with sparse data in the NIR to reconstruct the cosmography of our local supercluster, measure the growth-rate of structure and $H_0$, and test $\Lambda$CDM and alternative cosmological models. In the future, this can be extended to use optical data from the Rubin Observatory Legacy Survey of Space and Time (LSST) and NIR data from telescopes such as the Roman Space Telescope and James Webb Space Telescope.

This paper is organised as follows: in Section~\ref{sec:data}, we present the sample of SNe~Ia and the quality cuts used throughout this work, together with the \snoopy fits. In Section~\ref{sec:simulations}, we describe the simulations of NIR light curves. The comparison between the fits of the simulations and our reference sample is presented in Section~\ref{sec:analysis}, while the study of systematics is presented in Section~\ref{sec:systematics}. In Section~\ref{sec:nir_distances}, we estimate distances using the reference sample of SNe~Ia and compare them with those from the simulations. Finally, we summarise and conclude in Section~\ref{sec:conclusions}.

The scripts used throughout this work can be found online\footnote{\url{https://github.com/HOSTFLOWS/flows_sims}}.

\section{Data and Method}
\label{sec:data}

In this work, we use the Carnegie Supernova Project \citep[CSP;][]{Hamuy2006} sample as it is one of the most comprehensive samples of SNe~Ia with extensive \textit{uBgVriYJH} (optical to NIR) coverage and well-understood magnitude systems to date. The data from \cspi consists of three data releases (DRs) described in \cite[][DR1]{Contreras2010}, \cite[][DR2]{Stritzinger2011} and \cite[][DR3]{Krisciunas2017}, while the data from \cspii are described in \cite{Phillips2019} and \cite{Hsiao2019}. \cspii does not have a public DR to date, but one is in progress (Suntzeff et al., in prep.). We include all the 134 SNe~Ia from \cspi and 202 from \cspii (the cosmology sub-sample from \citealt{Phillips2019}). Thus, the CSP sample we use consists of a total of 336 SNe~Ia.

As CSP observations have used different filters and telescopes throughout the different campaigns, we apply S-corrections \cite[][]{Stritzinger2011} to work in a single magnitude system, simplifying the handling of data. This is specifically useful for those \cspi SNe with multiple \textit{V}, \textit{Y} and/or \J bands, and for combining SNe from \cspi and \cspii.

\subsection{Light-Curve Fitter}
\label{subsec:lc_fitter}

Currently, a small number of SNe~Ia light-curve fitters work in the NIR, such as \snoopy \citep{Burns2011}, \pisco \citep{Muller-Bravo2021} and \bayesn \citep{Mandel2022}. The difference between them is that \snoopy and \bayesn are template-based fitters (i.e. trained on well-sampled SNe~Ia) while \pisco uses a data-driven approach that relies on Gaussian Processes \citep[GPs;][]{Rasmussen2006}. However, for the aim of this work, as we need to reconstruct the NIR light curves with a few photometric points, \pisco is not an option. On the other hand, the difference between \snoopy and \bayesn is that the latter uses a probabilistic approach, constructing a hierarchical Bayesian model for the spectral energy distribution of SNe~Ia, therefore obviating the need of ad-hoc $K$-corrections (as is the case of \snoopy). However, \bayesn assumes a dust extinction law for SNe~Ia, so it does not have the same freedom as \snoopy. Additionally, \bayesn is not public, so we choose to use \snoopy with its latest version (\texttt{v2.6}) that has being trained on the updated NIR spectral templates from \cite{Hsiao2007}. However, \snoopy, like other light-curve fitters, heavily relies on optical data for accurate fitting.

\begin{figure}
	\includegraphics[width=\columnwidth]{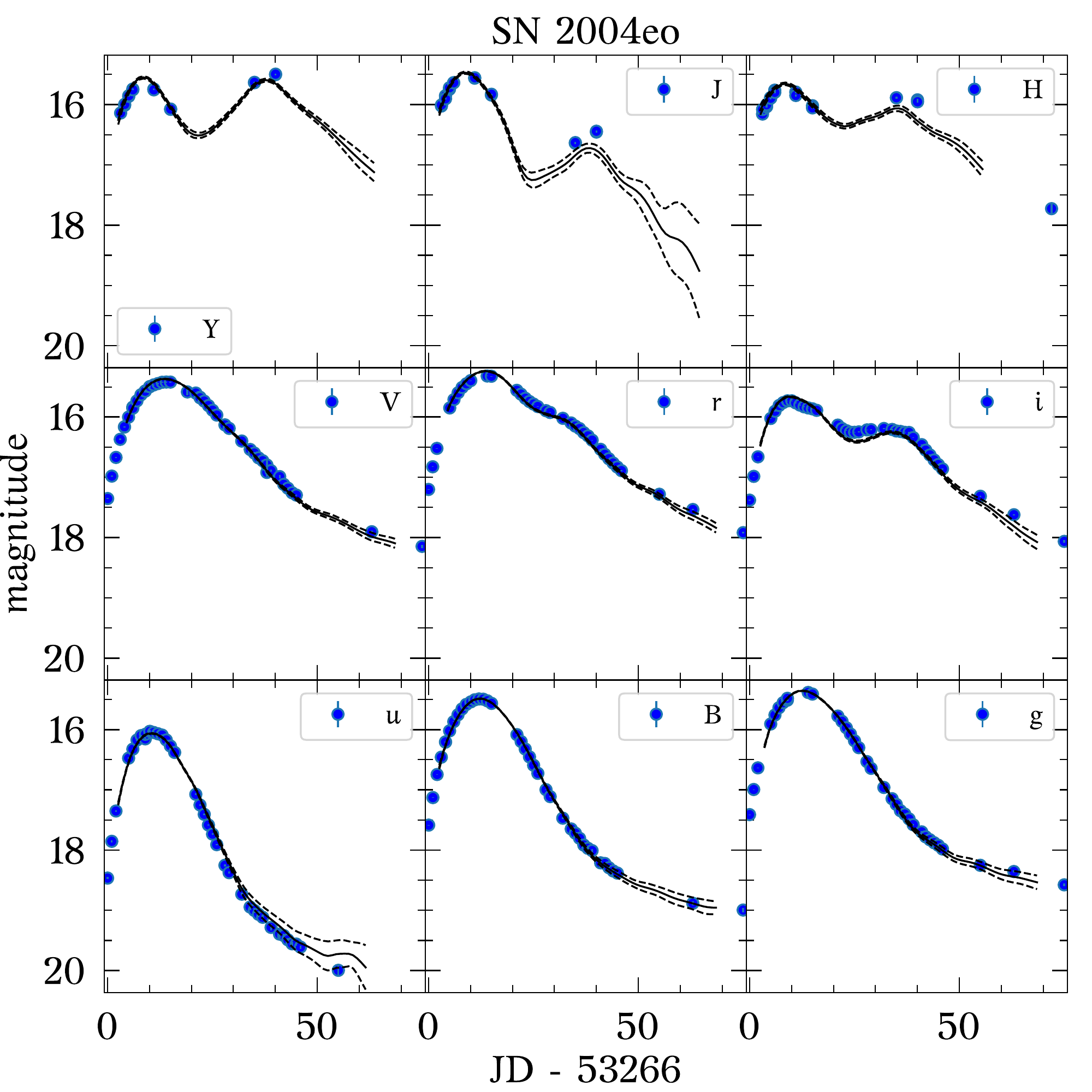}
    \caption{\snoopy fit of the multi-colour (\textit{uBgVriYJH}) light curves of SN~2004eo. The \maxmodel model is used throughout this work.}
    \label{fig:snoopy_fit}
\end{figure}

To fit the SNe Ia, we use \snoopy with the \maxmodel model as we require measurements of \Jmax and \Hmax. The resulting fits provide the following output parameters: \tmax, \sbv and $x_{\rm max}$, where \tmax is the time of \textit{B}-band peak magnitude, \sbv is the ``colour stretch\rq\rq\ parameter as defined in \citet[][]{Burns2014}\footnote{\sbv $=(T_{(B-V)_{\rm max}}-T_{\rm max})/30$\,days, where $T_{(B-V)_{\rm max}}$ is the time of maximum (reddest colour) in the $(B-V)$ colour curve} and $x_{\rm max}$ represents the peak magnitude in $x$-band, for each of the observed filters. An example fit for SN~2004eo is shown in Figure~\ref{fig:snoopy_fit}. Note that the multi-colour light-curve templates (optical to NIR) are driven by the values of \tmax and \sbv. For more information on why the \maxmodel model is used instead of other models, such as \EBVmodel, see Appendix~\ref{app:snoopy}. All magnitudes presented are in the CSP natural system and the reported uncertainties from \snoopy fits are statistical uncertainties only.

\subsection{Sample Cuts}
\label{subsec:sample_cuts}

As not all CSP SNe are useful for the purpose of this work, we proceed to apply some cuts to the initial sample. We only use SNe~Ia labelled as \lq\lq normal\rq\rq\ according to \cite{Krisciunas2017} and \cite{Ashall2020}. A current re-analysis of SNe~Ia performed by Phillips et al. (in prep.) has re-labelled 1991T-like SNe from the CSP sample. Thus, we use their definition hereon. This cut removes 34 of the initial 336 SNe~Ia: one 2003fg-like SN, four 1986G-like SNe, ten 1991T-like SNe, sixteen 1991bg-like SNe and three peculiar SNe (labelled as "..." in \citealt{Krisciunas2017}). 

We then proceed to remove any SN without \textit{g}, \textit{r}, \J or \H bands, as these are strictly required for our analysis: \textit{g} and \textit{r} being the bands used by ZTF while \J and \H being the NIR bands commonly available. \textit{Y} band is not included as catalogues of standard stars for this band are not available for the whole sky, which are needed for calibration. We note that 80 SNe do not have \textit{g} band (all from \cspii) while all of them have \textit{r} band. In this step we identify and remove 138 of the remaining 302 SNe.

The next cut requires a SN to have coverage of the optical peak as the estimation of \tmax is fundamental when fitting the light curves of SNe~Ia (Sections~\ref{subsec:ref_sample} \& \ref{sec:simulations}). For this, we need to have one or more photometric points at least two days before and two days after \tmax in \textit{B}, \textit{g}, \textit{V} or \textit{r} bands, providing an accurate estimation of the location of the peak. Almost all the SNe have the optical decline well covered, while some do not have pre-peak observations. Note that the SNe are fitted with \snoopy first to estimate \tmax. This cut removes 53 of the remaining 164 SNe. 

Finally, we require at least one photometric point before and after the time of \J-band peak (\Jmax) and \H-band peak (\Hmax), as precise measurements of \Jmax and \Hmax are needed. These constraints are less stringent compared to those in the optical as the light-curve fits are mainly driven by the optical bands. In addition, we require at least three photometric points, in each NIR band, to have a relatively good coverage of the light-curve peak. Note that the SNe are fitted with \snoopy first, with the same model as before, to estimate the locations of the NIR peaks. Here we remove 61 of the 111 remaining SNe, leaving us with a sample of 50 CSP SNe~Ia, comprising our reference sample.

\begin{figure}
	\includegraphics[width=\columnwidth]{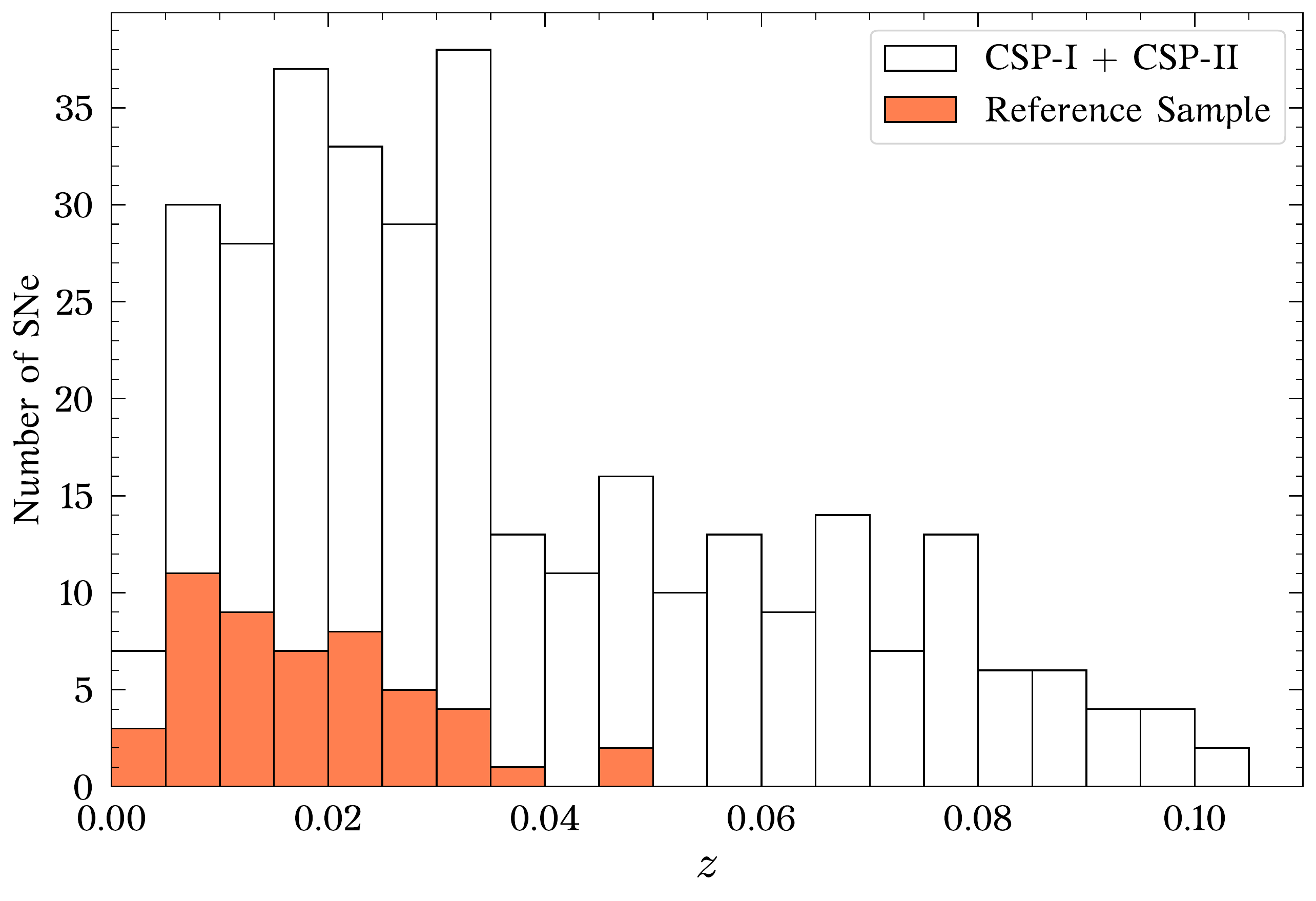}
	\includegraphics[width=\columnwidth]{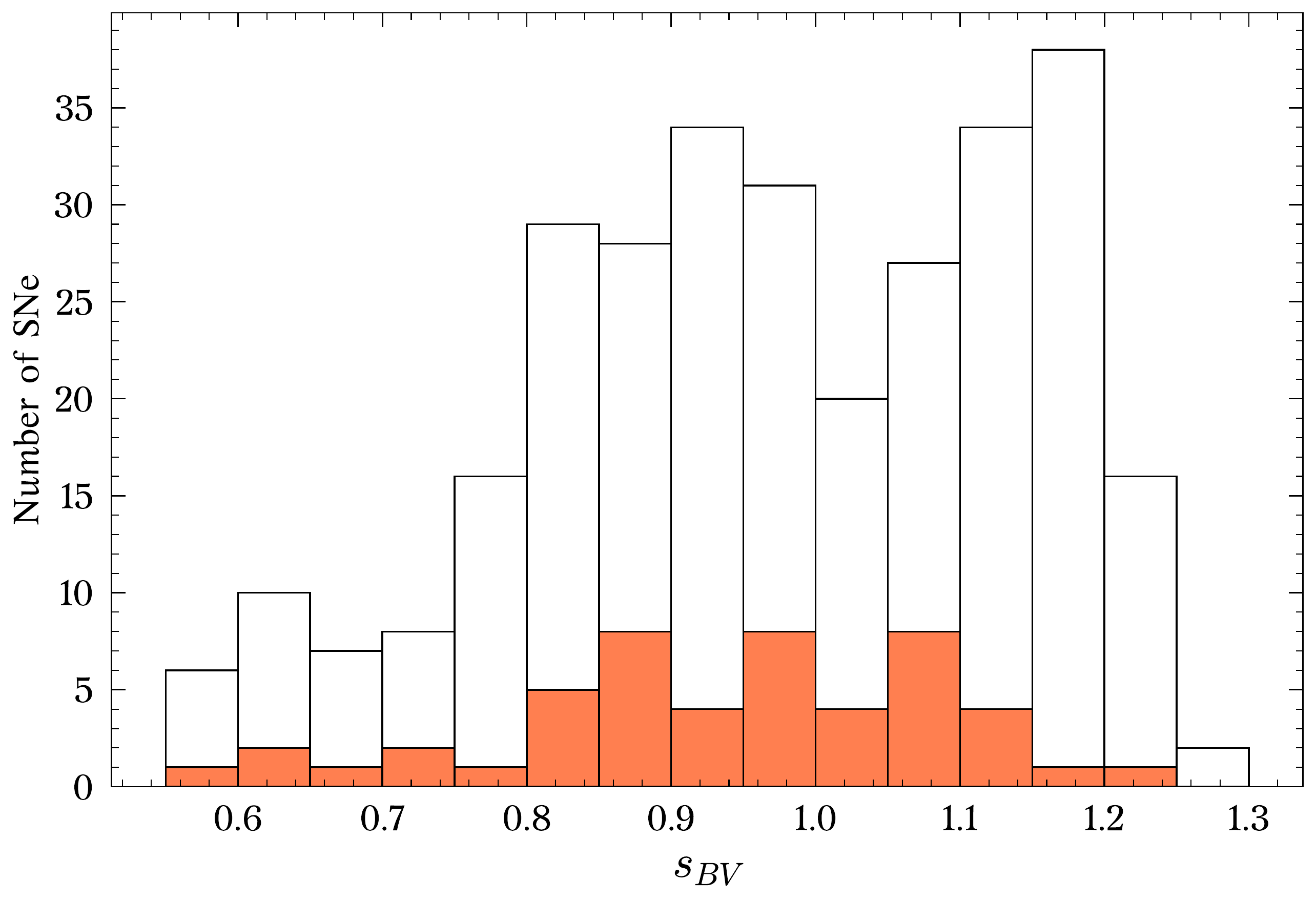}
    \caption{\textit{Top} panel: redshift distribution of \cspi $+$ \cspii (empty bars) and the 50 SNe~Ia comprising our reference sample (orange bars). \textit{Bottom} panel: \sbv distribution of SNe~Ia from \cspi $+$ \cspii (empty bars) labelled as ``normal\rq\rq\ and the reference sample (orange bars). The values of \sbv were obtained by fitting the SNe with \snoopy using the \maxmodel model. Only the ``normal\rq\rq\ SNe are shown in this case as \snoopy mainly provides templates for these objects and not other types.}
    \label{fig:z_dist}
\end{figure}

In the top panel of Figure~\ref{fig:z_dist}, we show the redshift distribution of the reference sample and the complete \cspi $+$ \cspii sample. One downside of the cuts applied is that all objects with $z\gtrsim0.05$ are removed. An object at $z=0.1$ is approximately $1.5$\,mag fainter than one at $z=0.05$, which translates into a poorer signal-to-noise ratio (S/N) of the observations, making the estimations of light-curve parameters less accurate. However, SNe at higher $z$ also have faster restframe cadences compared to those at lower $z$ (for the same observed cadence), although at $z\lesssim0.1$ there is but a small difference. The effects of cadence and S/N are later discussed in Sections~\ref{subsec:cadence}, \ref{subsec:snr} \& \ref{subsec:nir_snr}.

In the bottom panel of Figure~\ref{fig:z_dist}, we show the \sbv distribution of the reference sample and the ``normal\rq\rq\ SNe~Ia from the \cspi $+$ \cspii sample. Both distributions span approximately the same range in \sbv, with no major differences apart from the number of objects.

\begin{center}
\begin{table}
\caption{Number of CSP SNe Ia discarded by the cuts outlined in Section~\ref{subsec:sample_cuts}.}
\centering
\begin{tabular}{ccc}
\hline\hline

Cut & Removed & Remaining  \\ 
\hline
\hline
Initial sample          &               & 336 (134|202)  \\ 
\hline

"normal" type           & 34 (21|13)     & 302 (113|189)  \\
\textit{grJH} bands     & 138 (21|117)  & 164 (92|72)  \\
Optical peak coverage   & 53 (30|23)    & 111 (62|49)  \\
NIR peak coverage       & 61 (21|40)    & 50  (41|9) \\
\hline

\end{tabular}
\begin{tablenotes}
 \item \textbf{Notes.} The values in parentheses are the number of SNe Ia from \cspi (left|) and \cspii (|right).
\end{tablenotes}
\label{tab:cuts}
\end{table}
\end{center}

The cuts from this section are summarised in Table~\ref{tab:cuts}, where the numbers are also split by survey (\cspi and \cspii). Tighter constraints could be used, although this would further reduce the size of our sample, reducing statistics as well, and change the distribution of light-curve parameters.

\subsection{Reference Sample}
\label{subsec:ref_sample}

\begin{figure}
	\includegraphics[width=\columnwidth]{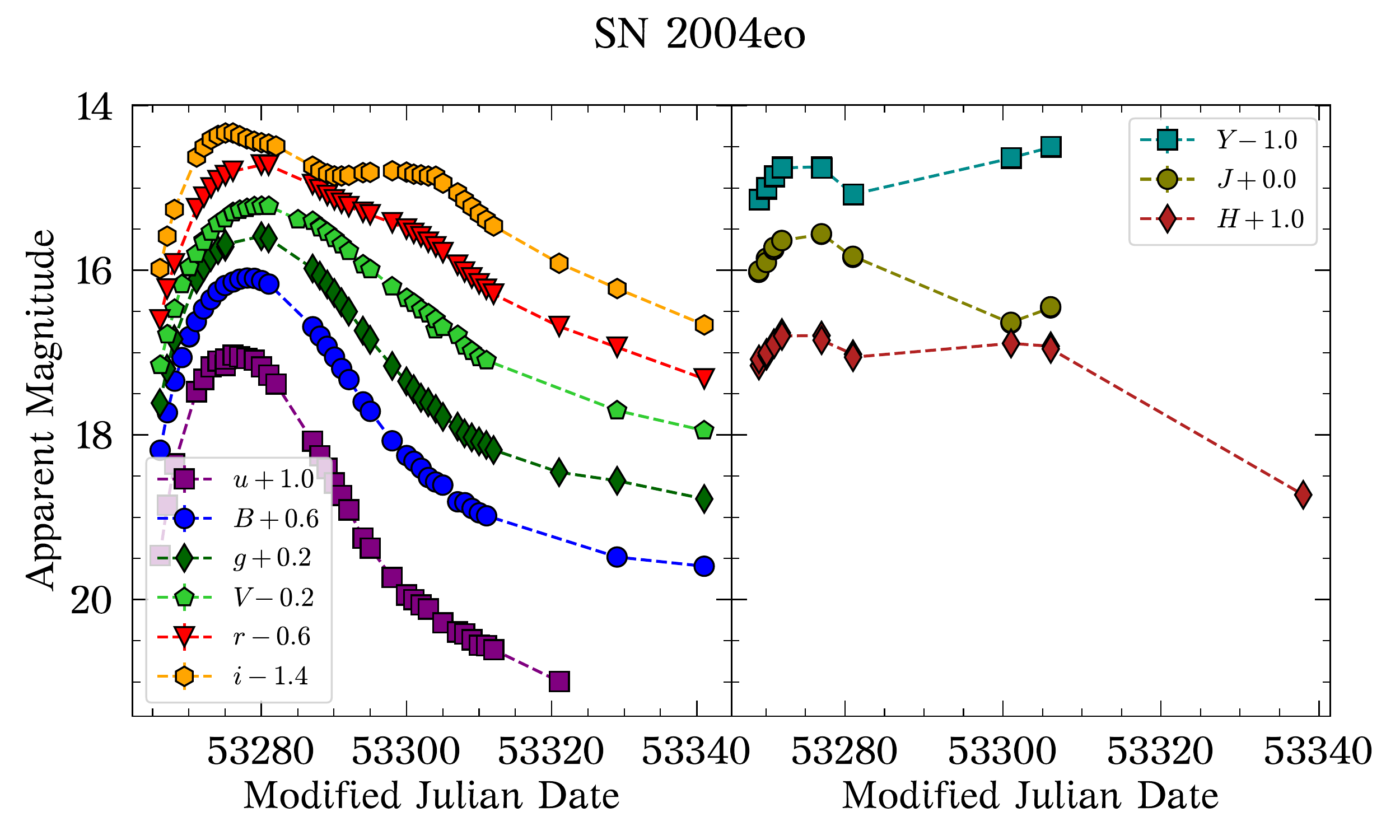}
	\includegraphics[width=\columnwidth]{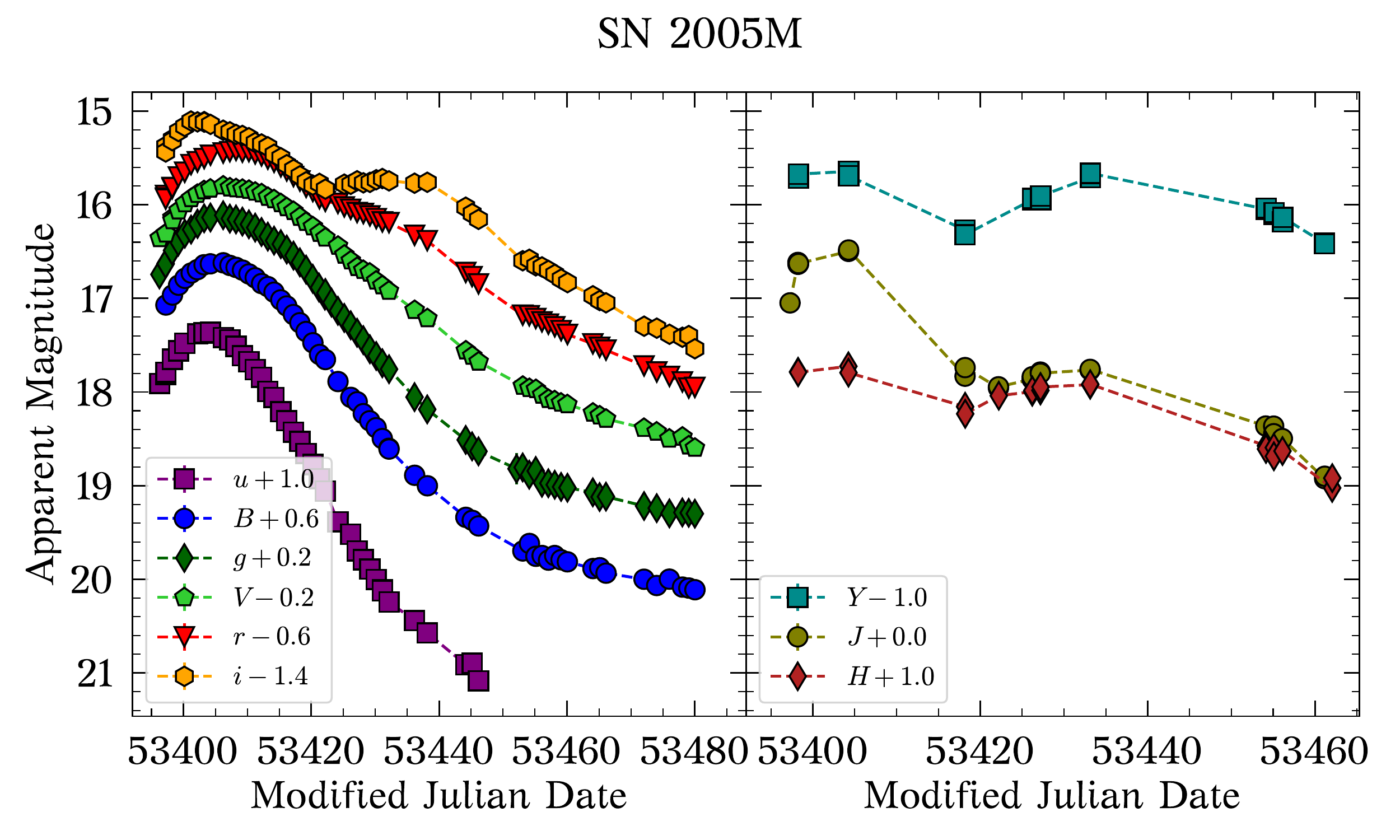}
	\includegraphics[width=\columnwidth]{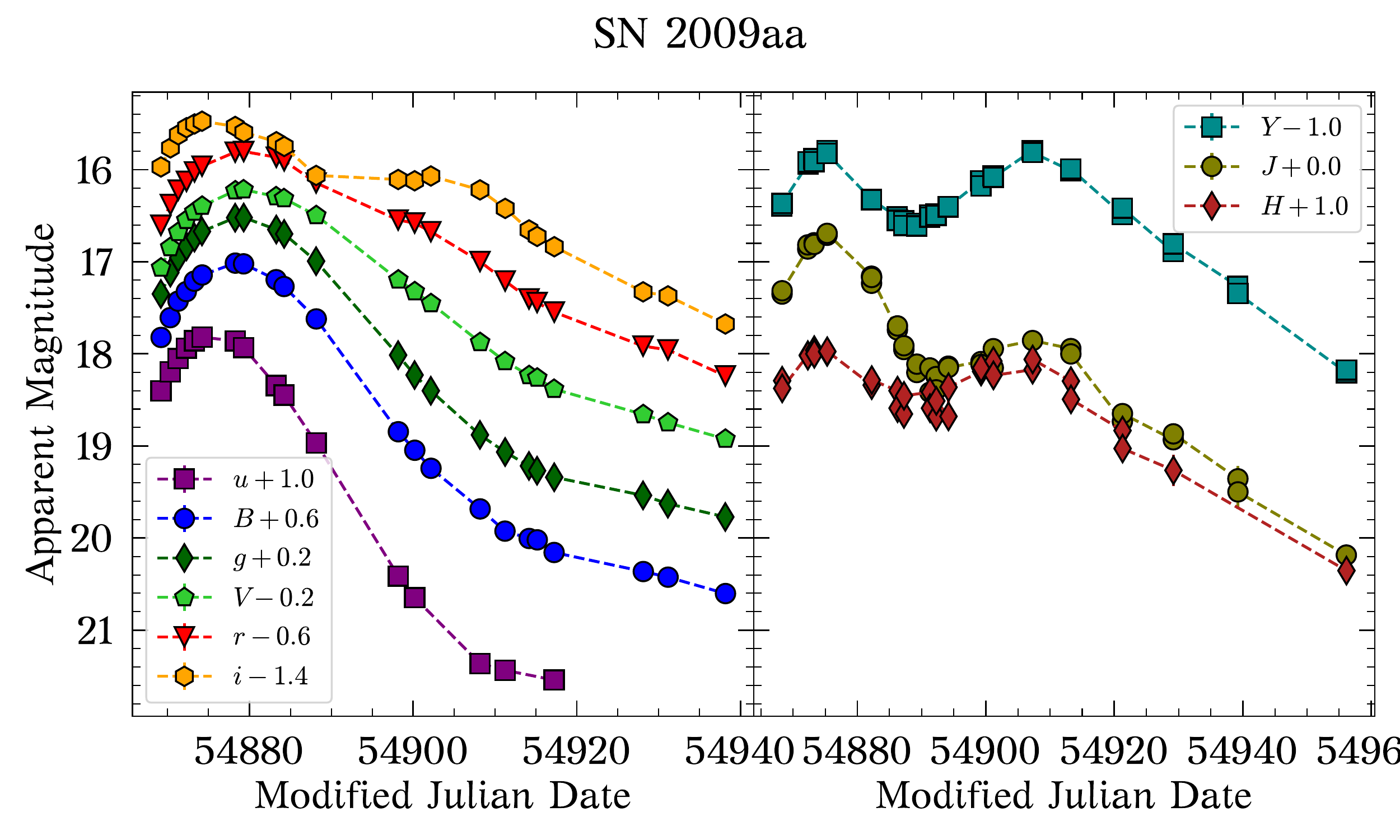}
    \caption{Multi-colour light curves of SNe 2004eo (\textit{top}), 2005M (\textit{middle}) and 2009aa (\textit{bottom}) from \cspi.}
    \label{fig:example_csp1}
\end{figure}

\begin{figure}
	\includegraphics[width=\columnwidth]{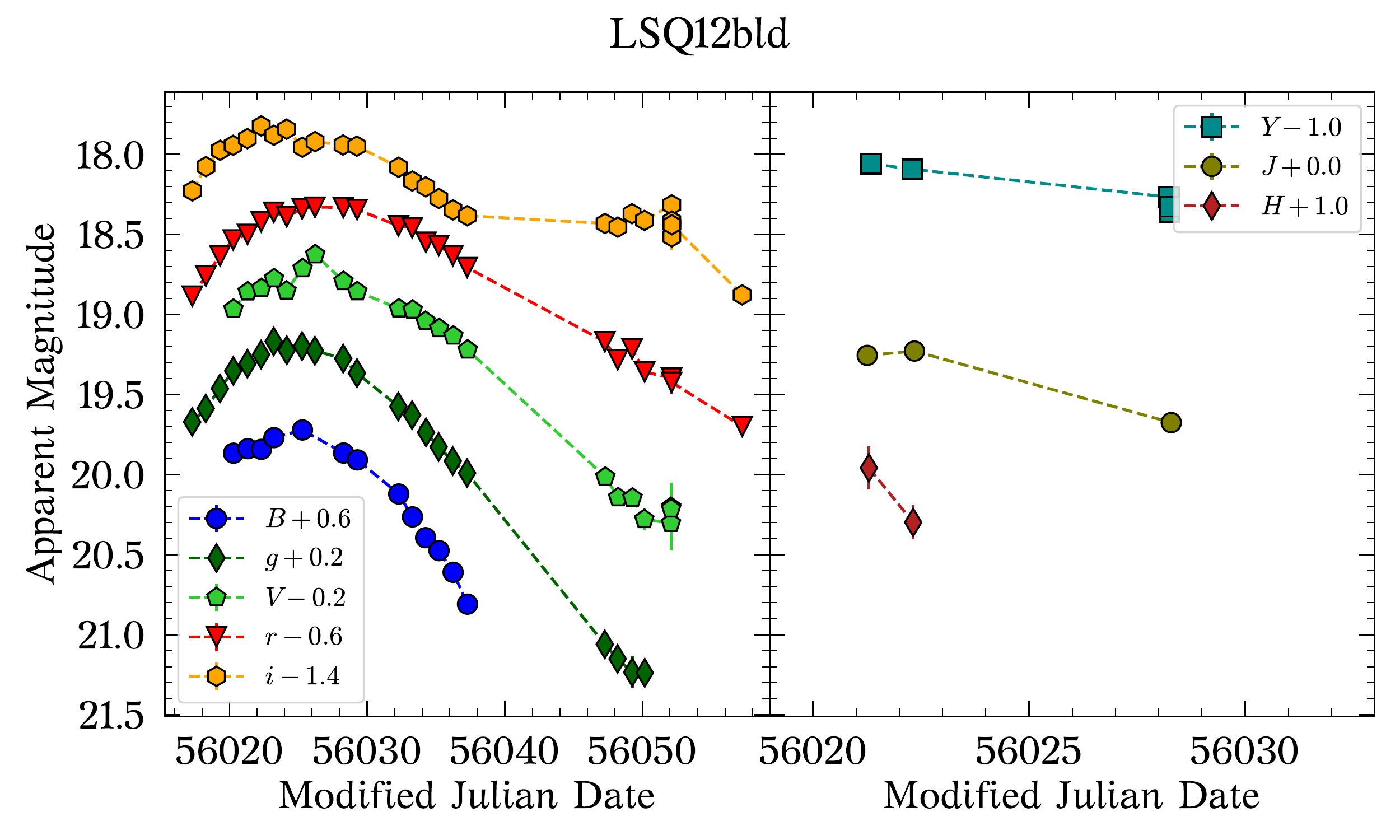}
	\includegraphics[width=\columnwidth]{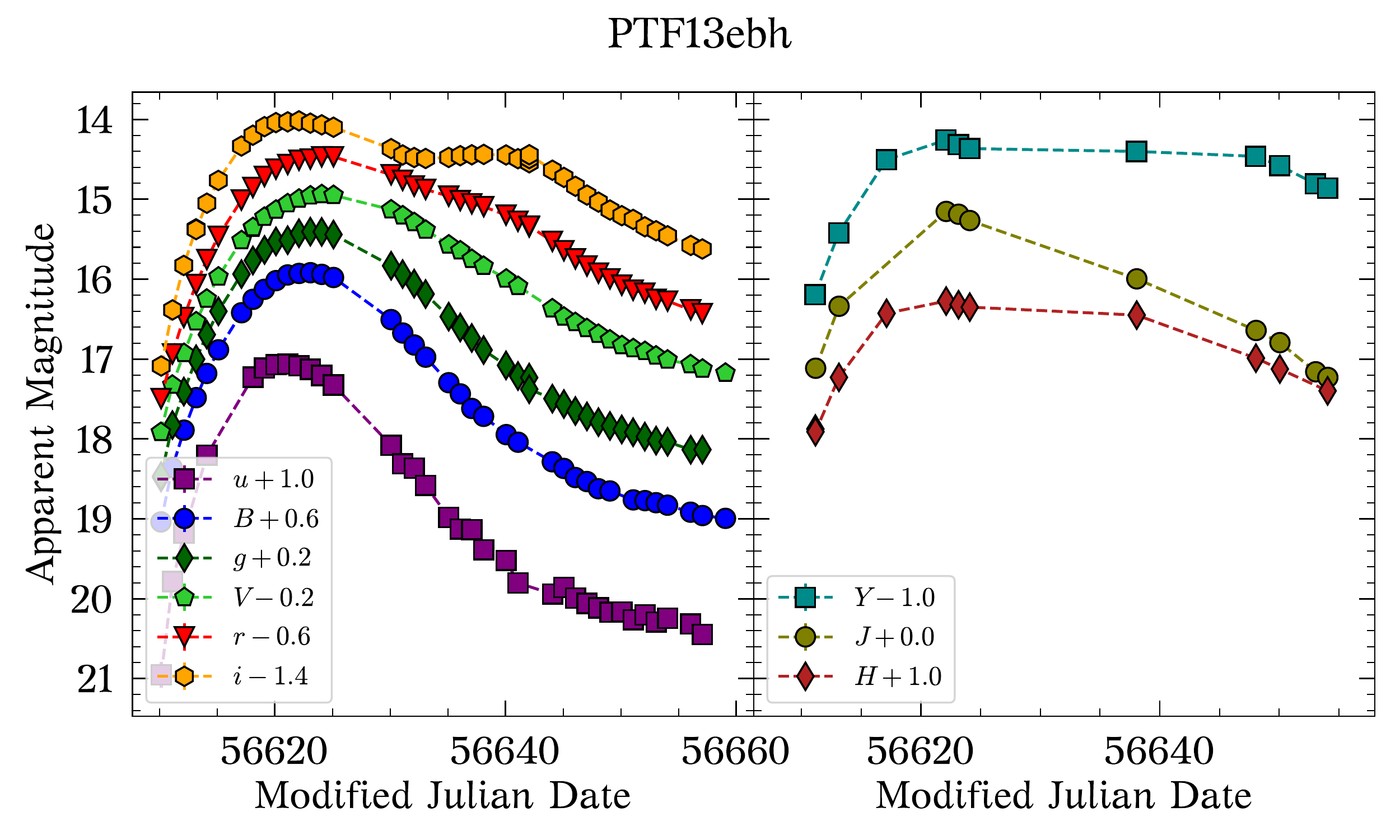}
	\includegraphics[width=\columnwidth]{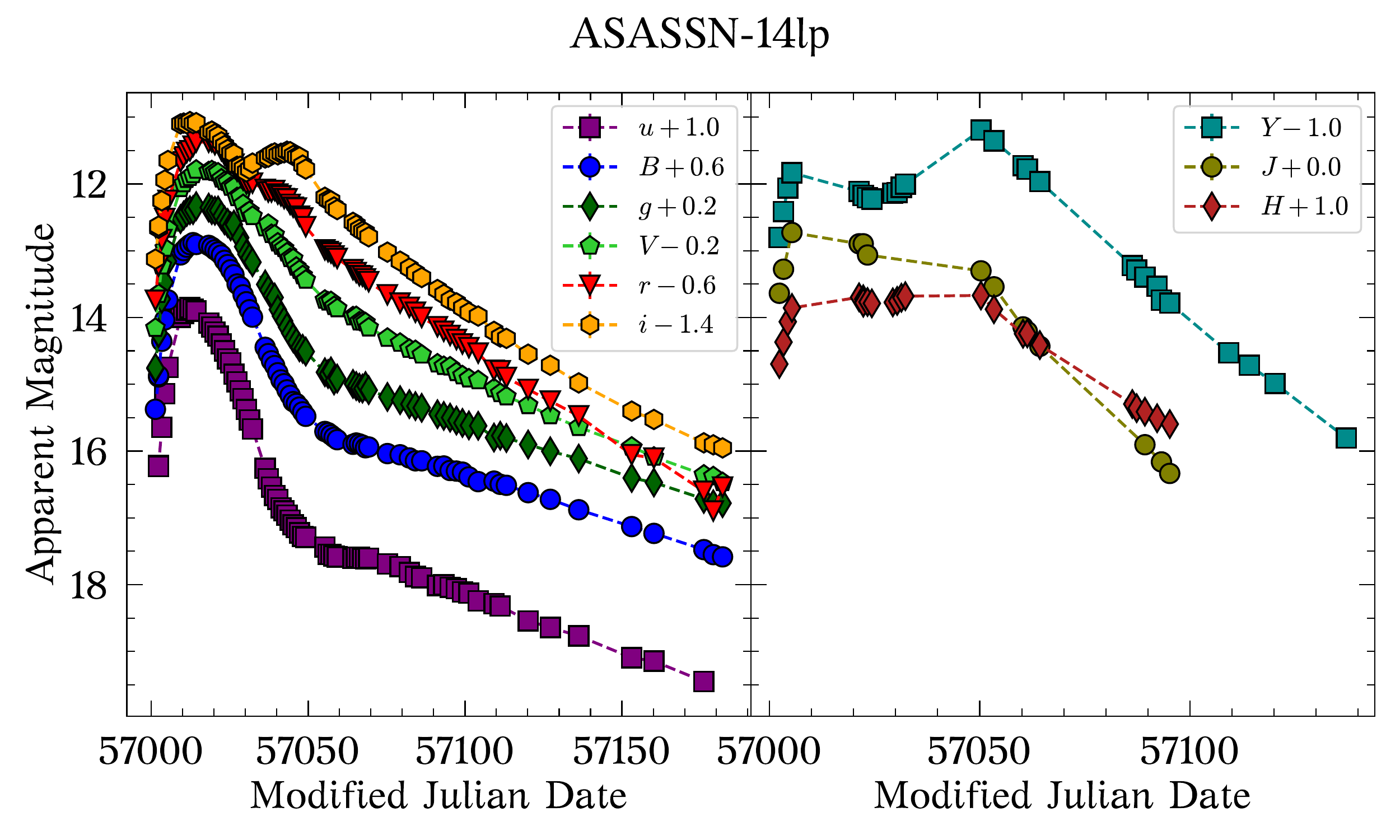}
    \caption{Multi-colour light curves of SNe LSQ12bld (\textit{top}), PTF13ebh (\textit{middle}; \citealt{Hsiao2015}) and ASASSN-14lp (\textit{bottom}; \citealt{Shappee2016}) from \cspii.}
    \label{fig:example_csp2}
\end{figure}

Although the data quality of \cspi and \cspii are almost identical, the data coverage in the NIR bands is not the same. The NIR light curves of SNe~Ia from \cspi are on average better populated than those from \cspii. Examples of SNe from both surveys are shown in Figures~\ref{fig:example_csp1} \& \ref{fig:example_csp2}. This sample of 50 SNe~Ia are later used as reference for the simulations in Section~\ref{sec:simulations}.

We create three different sets of reference values according to the bands used for the fits: (i) all bands; (ii) \textit{grJH}; and (iii) \textit{gr} (see Table~\ref{tab:fits}). Note that sets (ii) and (iii) are mainly used as control sets to test whether there are any discrepancies in the output parameters (e.g., \tmax) compared to using all bands (see Section~\ref{sec:analysis}).

\section{Simulations}
\label{sec:simulations}

Given that we want to replicate what real observations would be (i.e. optical bands well covered with few NIR data points), the simulations consist of taking the complete \textit{gr}-band light curves, plus $n$ epochs of coeval \J- and \H-band photometric points, for $n = 1, 2$ and $3$. Note that we are sampling from the available photometry (this is what we call simulations hereon). Combinations without repetition of the \textit{JH}-band photometry are used for this:

\begin{equation}
    C_{n}(N)=\left(\begin{array}{l} N\\n\end{array}\right)=\frac{N !}{n !(N-n) !},
\end{equation}

\noindent where $C$ is the total number of combinations and $N$ is the total amount of $J$/$H$ epochs, for each band individually. For instance, a SN with 10 epochs of coeval \J- and \H-band photometric points would have $C_{1}(10) = 10$, $C_{2}(10) = 45$ and $C_{3}(10) = 120$ simulations for 1, 2 and 3 $J$/$H$ epochs, respectively.

\begin{figure}
	\includegraphics[width=\columnwidth]{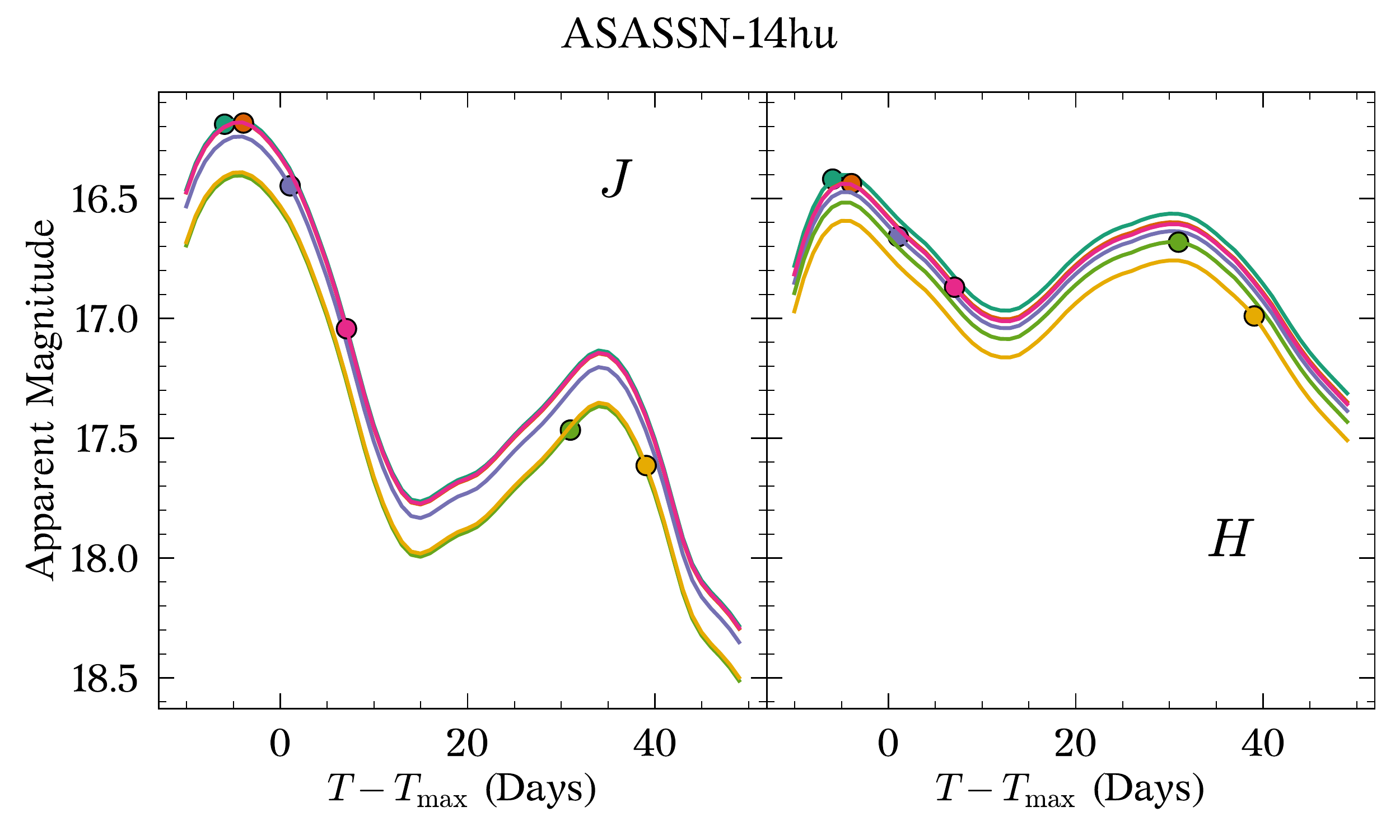}
    \caption{Fits of single-epoch \textit{JH} simulations for ASASSN-14hu. Each colour represents a different simulation and fit for the NIR photometry of the SN.}
    \label{fig:sim_example}
\end{figure}

The resulting simulations are then fit, using \snoopy as described in Section~\ref{subsec:ref_sample}, and the output parameters saved for a later comparison (Section~\ref{sec:analysis}). In Figure~\ref{fig:sim_example}, we show an example of simulations with $n = 1$ for ASASSN-14hu, with the respective fits for each simulation in \J and \H bands. As can be seen from the fits, the values of \Jmax and \Hmax highly depend on the epochs of the photometry. Note that, as we are dealing with few \J and \H photometric epochs, a proper estimation of the NIR peaks is, in principle, not possible, so we measure the time with respect to \tmax, which can be accurately obtained from the complete optical \textit{gr}-band light curves.

Simulations are also performed for \J and \H bands separately to test whether having only one NIR band produces similar results compared to having two. To summarise, we have simulations with \textit{grJ}, \textit{grH} and \textit{grJH} bands (see Table~\ref{tab:fits}), each with 1, 2 and 3 NIR epochs. The different sets of fits for the reference sample and simulations are summarised in Table~\ref{tab:fits}.

\section{Analysis}
\label{sec:analysis}

Assuming that SNe Ia are standard candles in the NIR, peak NIR magnitudes can be directly used to estimate distances without further corrections (e.g., from stretch or colour). Thus, our main interest is to see how well we can measure \Jmax and \Hmax using just a few NIR photometric points. However, as can be seen in Figure~\ref{fig:sim_example}, the retrieved peak magnitudes highly depend on the time of the observations with respect to \tmax. We therefore define three different metrics to understand what type of observations are best at obtaining accurate \Jmax and \Hmax measurements:

\begin{enumerate}[(i)]
    \item time of the closest $J$/$H$ epoch with respect to \tmax;
    \item mean time of the $J$/$H$ epochs with respect to \tmax;
    \item difference between the earliest and latest (i.e. range) $J$/$H$ epochs.
\end{enumerate}

\noindent Note that in the case of the simulations with only one epoch ($n = 1$), metrics (i) and (ii) are the same, while (iii) is not calculated. Also note that these metrics are defined in the restframe, i.e. epochs are corrected for time dilation using the SN redshift.

\begin{center}
\begin{table}
\caption{Bands used for the fits of the reference sample (Section~\ref{subsec:ref_sample}) and simulations (Section~\ref{sec:simulations}).}
\centering

\begin{tabular}{c|c}
\hline
sample & bands \\ 
\hline

reference & all bands, \textit{grJH}, \textit{gr} \\
simulations &  \textit{grJH}, \textit{grJ}, \textit{grH}\\
\hline
\end{tabular}

\label{tab:fits}
\end{table}
\end{center}

\begin{figure}
	\includegraphics[width=\columnwidth]{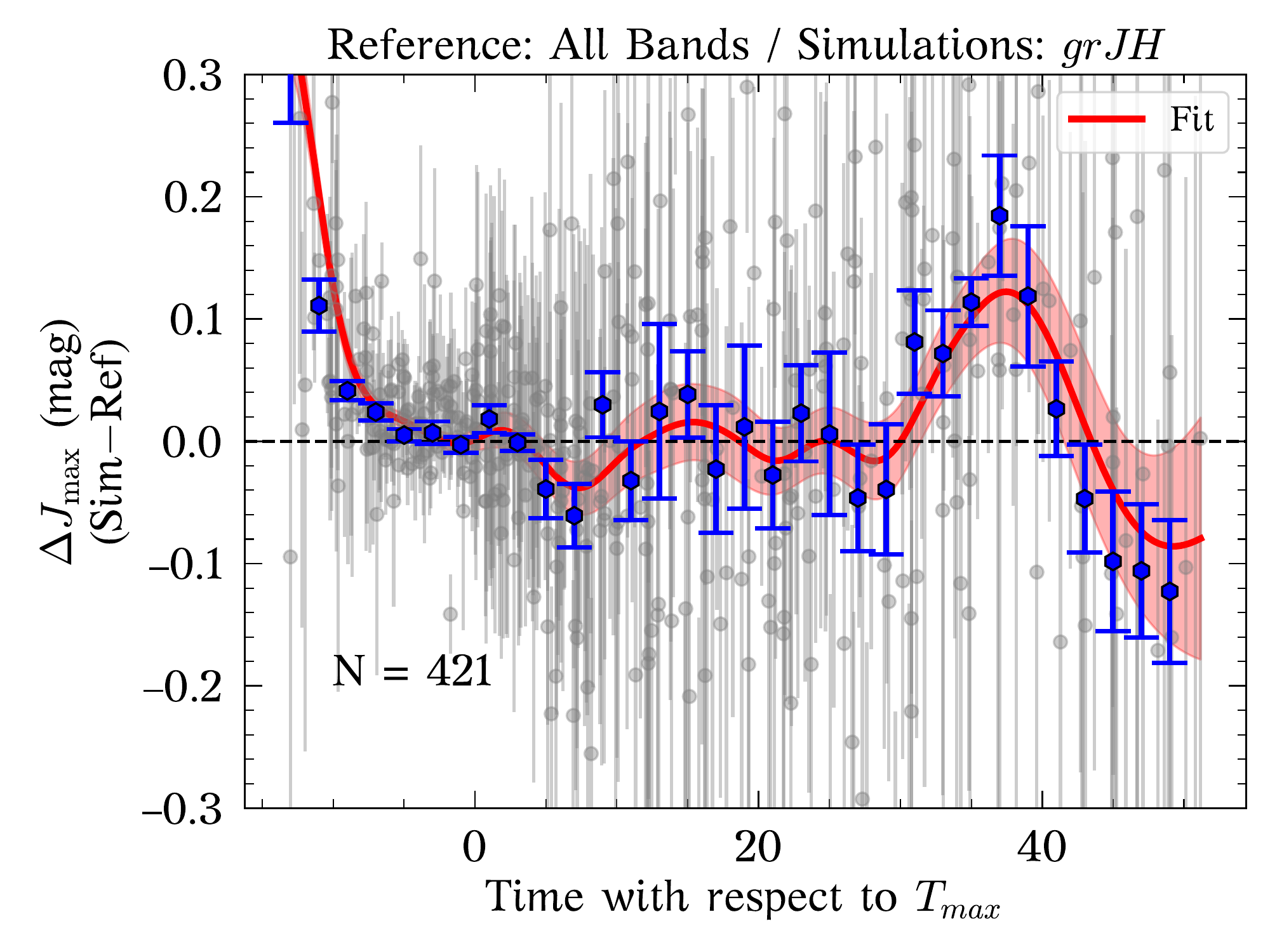}
	\includegraphics[width=\columnwidth]{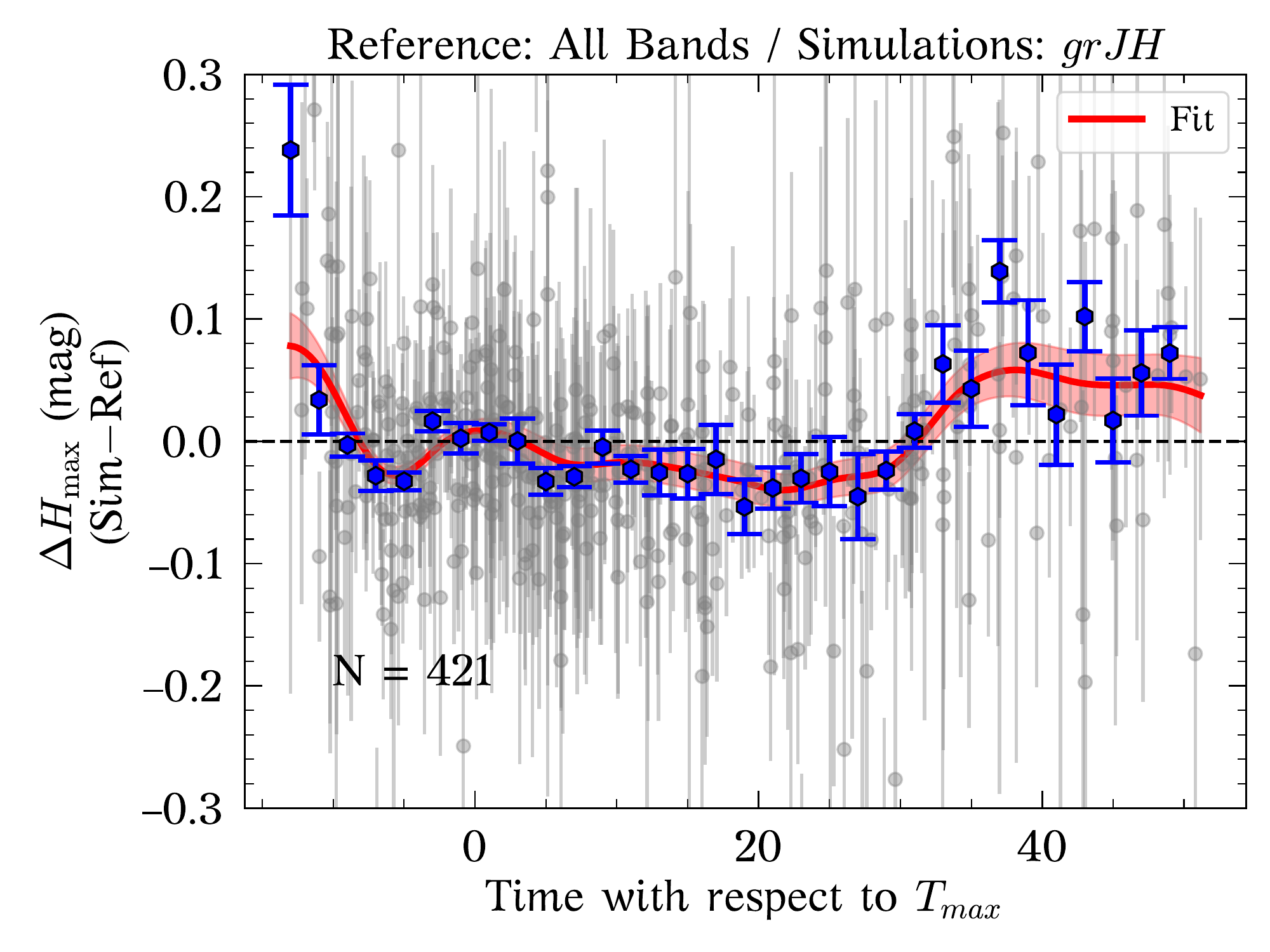}
    \caption{\Jmax (\textit{top} panel) and \Hmax (\textit{bottom} panel) residuals, between simulations with coeval coeval \J- and \H-band epoch and reference value. The weighted mean and uncertainty on the weighted mean in bins of 2 days are shown in blue. A ``correction snake\rq\rq\ and its uncertainty are calculated by fitting the residuals with GPs (red line and shaded region). N is the total number of simulations.}
    \label{fig:residuals}
\end{figure}

In Figure~\ref{fig:residuals}, we show \Jmax and \Hmax residuals between the simulations and the reference sample (using all bands) as a function of metric (i), for simulations with \textit{grJH} bands and $n = 1$. From this comparison, we see that the scatter in the residuals tends to be smaller around \tmax, with the smallest scatter before \tmax, and increases at later epochs. In general, the scatter is smaller for the \H band compared to the \J band. This is consistent with what has been found in other works \citep[e.g.][]{Dhawan2015}. Additionally, we note that offsets in \Jmax and \Hmax tend to be larger where the NIR light-curve templates have a larger gradient/slope (see Figure~\ref{fig:sim_example} and Section~\ref{sec:systematics} for further discussion). The weighted mean and the uncertainty on the weighted mean of these residuals are summarised in Table~\ref{tab:residuals}. Given that the uncertainties in light-curve parameters from the reference sample and simulations are correlated (both use the same photometry), hereon we only use those from the latter for the statistics.

As we are aiming to reduce the scatter, ideally, we would need data around \tmax. Unfortunately, as it is hard to obtain photometric data at specific epochs due to different constraints (weather, time allocation, etc.), we have to look for a time window with a large-enough range. We noticed that a time range between $-$5 to 15 days with respect to \tmax possess a relatively low scatter in \Jmax and \Hmax. The weighted mean ($\Delta$) and weighted standard deviation ($\sigma$) of the residuals in this time window are $\Delta = 0.004$\,mag and $\sigma = 0.047$\,mag, and $\Delta = -0.006$\,mag and $\sigma = 0.053$\,mag, for \Jmax and \Hmax, respectively.

\begin{figure}
	\includegraphics[width=\columnwidth]{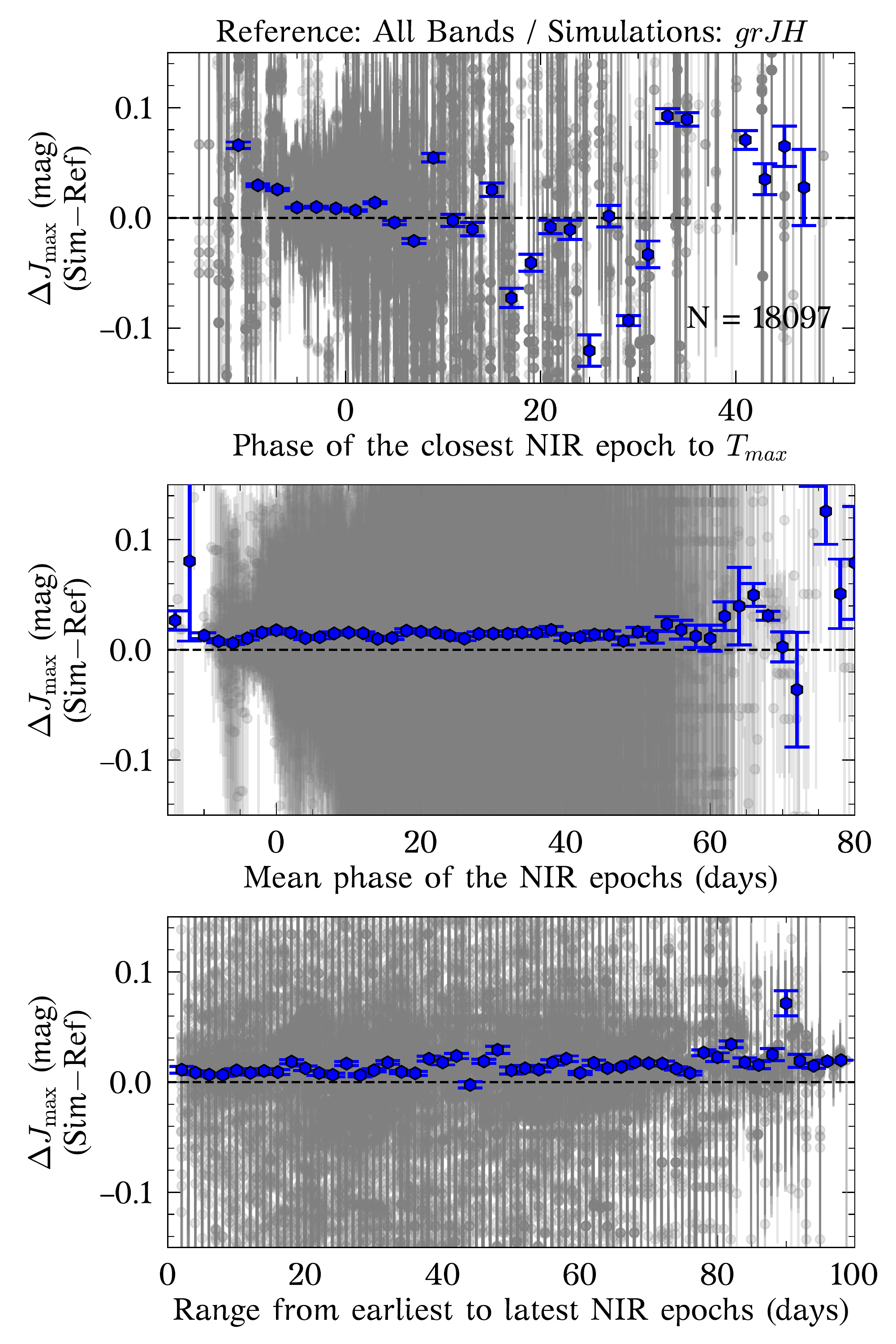}
    \caption{\Jmax residuals between simulations with three coeval \J- and \H-band epochs and reference sample as a function of metrics (i), (ii) and (iii) (\textit{top}, \textit{middle} and \textit{bottom} panels, respectively). The weighted mean and uncertainty on the weighted mean in bins of two days are shown in blue for each of the panels. The residuals in \Jmax do not vary much as a function of metrics (ii) and (iii), which means that metric (i) is the most relevant for estimating \Jmax. The metrics are described in Section~\ref{sec:analysis}. N is the total number of simulations. The global offsets in the middle and bottom panels are driven by those simulations with phases $\lesssim-5$\,days and $\gtrsim30$\,days from the top panel.}
    \label{fig:residuals_m123}
\end{figure}

In the case of the simulations with \textit{grJ} or \textit{grH} bands, similar results are obtained, meaning that a single NIR band is as good as two NIR bands (see Figure~\ref{fig:residuals_app} in the Appendix). When comparing with the simulations with $n = 2$ and $n = 3$, the scatter in the residuals is reduced. For $n = 2$, we find a scatter in \Jmax of $0.039$\,mag, while for $n = 3$, we find $0.030$\,mag. In the case of \Hmax, the scatter gets reduced to $0.043$\,mag and $0.038$\,mag, for $n = 2$ and $n = 3$, respectively. Although the offsets change for $n = 2, 3$, they still remain relatively small ($\lesssim0.01$\,mag). We also found that metrics (ii) and (iii) are not as relevant as metric (i) (see Figure~\ref{fig:residuals_m123} for simulations with $n = 3$). In other words, the most important point is to have a NIR photometric epoch close to \tmax. Having more photometric points helps reducing the scatter, although they are less essential for an accurate estimation of \Jmax or \Hmax.

\begin{center}
\setlength{\tabcolsep}{2pt}

\begin{table*}
\caption{Weighted mean ($\Delta$) and uncertainty on the weighed mean ($\sigma$) of \Jmax and \Hmax residuals between the reference sample (Section~\ref{subsec:ref_sample}) and simulations (Section~\ref{sec:simulations}).}
\centering

\begin{tabular}{c|cccccccccccccc}
\hline
Phase range & [$-$10,$-$8] & [$-$8,$-$6] & [$-$6,$-$4] & [$-$4,$-$2] & [$-$2,0] & [0,2] & [2,4] & [4,6] & [6,8] & [8,10] & [10,12] & [12,14] & [14,16] & [16,18] \\
\hline
\hline
$\Delta_{JH}$($J_{\rm max}$) & 0.054 & 0.024 & 0.019 & 0.005 & $-$0.005 & 0.012 & 0.009 & $-$0.028 & $-$0.042 & $-$0.033 & $-$0.008 & 0.032 & 0.041 & 0.007 \\
$\sigma_{JH}$($J_{\rm max}$) & 0.011 & 0.007 & 0.006 & 0.005 & 0.009 & 0.010 & 0.008 & 0.014 & 0.027 & 0.027 & 0.036 & 0.047 & 0.041 & 0.050 \\
\hline
$\Delta_{JH}$($H_{\rm max}$) & $-$0.012 & $-$0.033 & $-$0.014 & $-$0.009 & 0.005 & 0.007 & $-$0.005 & $-$0.007 & $-$0.024 & $-$0.018 & $-$0.012 & $-$0.027 & $-$0.022 & $-$0.020 \\
$\sigma_{JH}$($H_{\rm max}$) & 0.013 & 0.010 & 0.011 & 0.009 & 0.011 & 0.007 & 0.009 & 0.018 & 0.008 & 0.011 & 0.010 & 0.014 & 0.019 & 0.026 \\
\hline
$\Delta_{J}$($J_{\rm max}$) & 0.086 & 0.034 & 0.019 & 0.005 & 0.008 & 0.016 & 0.004 & $-$0.020 & $-$0.047 & $-$0.034 & $-$0.014 & 0.045 & 0.046 & 0.014 \\
$\sigma_{J}$($J_{\rm max}$) & 0.014 & 0.006 & 0.005 & 0.004 & 0.007 & 0.009 & 0.008 & 0.014 & 0.021 & 0.024 & 0.025 & 0.039 & 0.031 & 0.039 \\
\hline
$\Delta_{H}$($H_{\rm max}$) & $-$0.003 & $-$0.017 & 0.007 & 0.003 & 0.016 & 0.030 & 0.014 & 0.007 & $-$0.008 & $-$0.004 & 0.011 & $-$0.015 & $-$0.003 & $-$0.007 \\
$\sigma_{H}$($H_{\rm max}$) & 0.012 & 0.008 & 0.008 & 0.008 & 0.009 & 0.006 & 0.009 & 0.013 & 0.008 & 0.010 & 0.009 & 0.013 & 0.014 & 0.017 \\
\hline

\end{tabular}
\begin{tablenotes}
 \item \textbf{Notes.} The data from the first couple of rows are represented by blue circles with error bars in Figure~\ref{fig:residuals}. The NIR bands used for the light-curve simulations are given as subscripts. For instance, $\Delta_{JH}$(H$_{max}$) is the weighted mean of \Hmax residuals using \textit{J}- and \textit{H}-band light curves.
\end{tablenotes}

\label{tab:residuals}
\end{table*}
\end{center}

When comparing other light-curve parameters, such as \gmax, \rmax, \sbv and \tmax, between the reference sample and simulations, we found negligible differences ($\lesssim0.02$\,mag) for the first three parameters and $<0.1$\,day for the last. {These very small differences are produced by the NIR data. Although the fits are mainly driven by the optical light curves, the NIR also affects the estimation of \tmax, and therefore, the other light-curve parameters.} When using \textit{grJH} bands and \textit{gr} bands as reference, we found similar results as when using all bands as reference (note that for \textit{gr} bands only, \Jmax and \Hmax cannot be calculated).

\section{Data-Quality Systematics}
\label{sec:systematics}

As was previously mentioned, the light-curve fits heavily rely on having a proper estimation of \tmax. Any uncertainty in this parameter is propagated to others, such as the NIR peak magnitude. It follows that having small uncertainties and fast cadence is ideal, particularly in the optical bands. However, this is not always possible. In this section, we test how cadence and S/N affect the estimations of \Jmax and \Hmax.

\subsection{Cadence}
\label{subsec:cadence}

To measure the effect of cadence, we started by taking the \textit{gr}-band light curves of the reference sample and simulated different cadences by taking every two observations in each light curve, every three, every four and so on. We make sure to ``move\rq\rq\ the first observation taken so we have different starting points until all epochs before \tmax have been used (the reader should not confuse these simulations with those from Section~\ref{sec:simulations}). Note that the observer-frame cadences across the optical filters are very similar for a single SN as CSP observations were taken with multiple filters during the same nights (see Figures~\ref{fig:example_csp1} \& \ref{fig:example_csp2}). Our reference sample has an average cadence of $\sim3.1$\,days in \textit{gr}, while the simulations have cadences $>3.1$\,days.

\begin{figure}
	\includegraphics[width=\columnwidth]{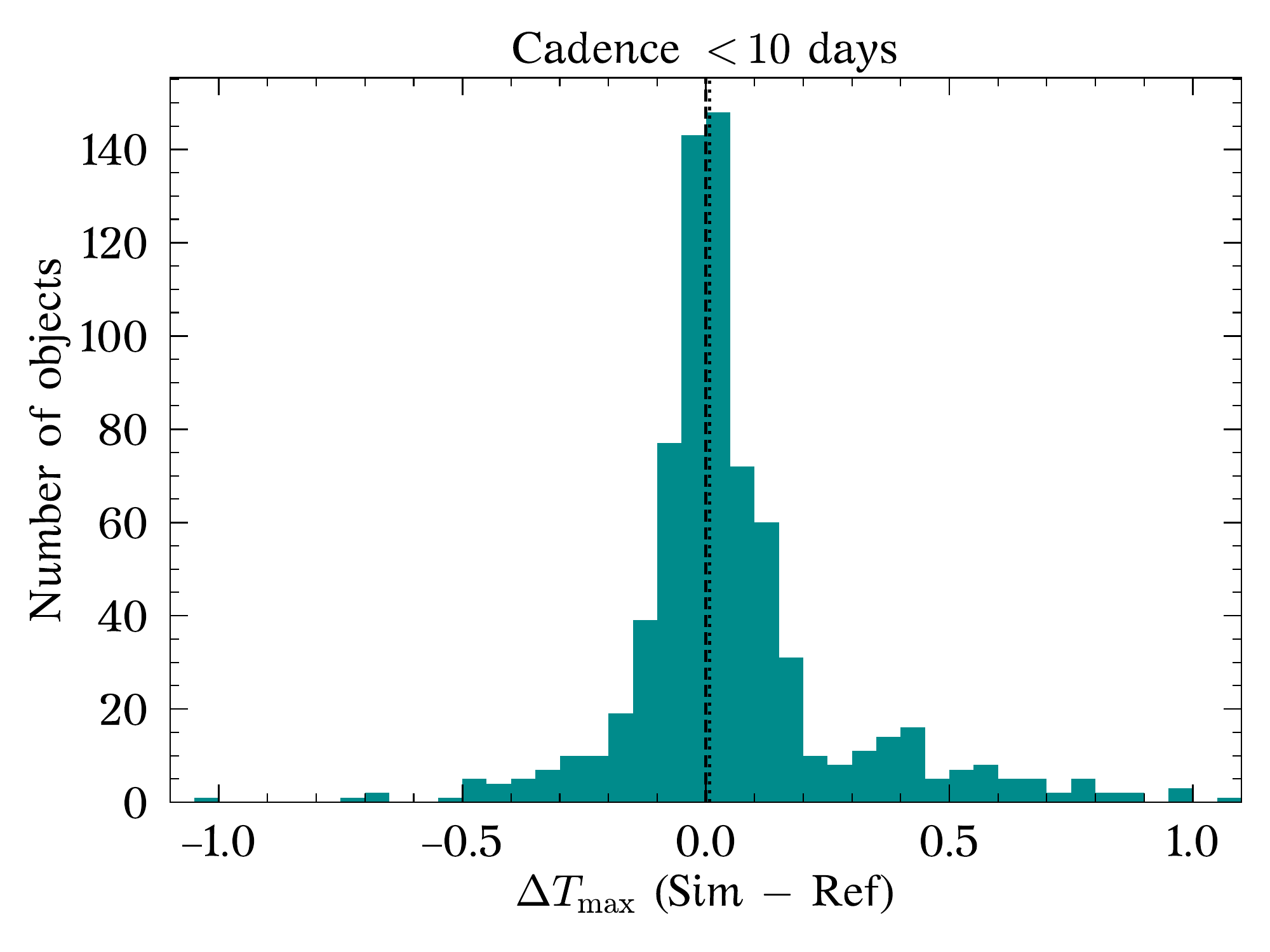}
	\includegraphics[width=\columnwidth]{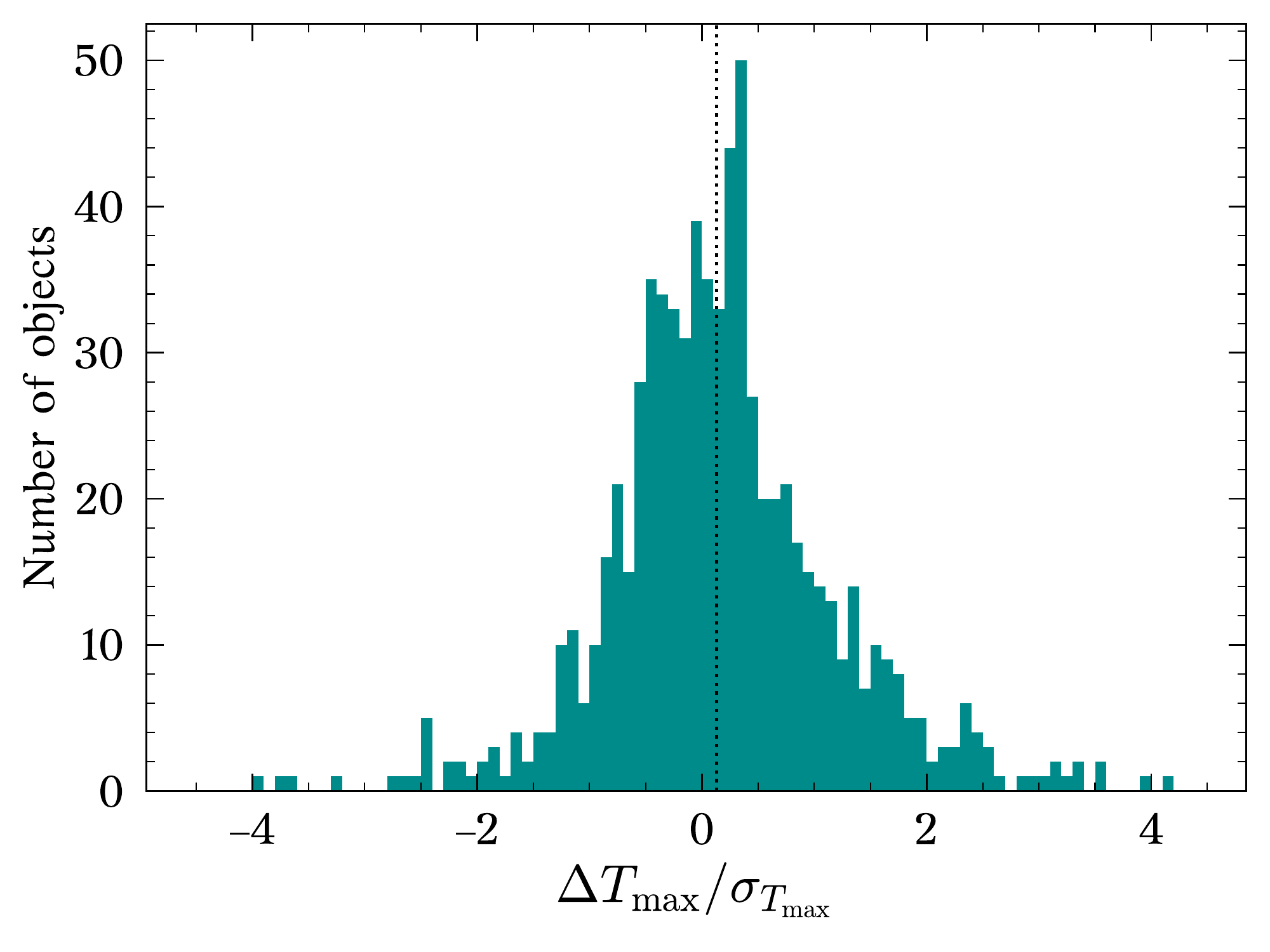}
    \caption{\textit{Top} panel: difference in \tmax between the cadence simulations (Section~\ref{subsec:cadence}) and reference sample (\dtmax). All simulations with restframe cadences $<10$\,days are considered here. The weighted average \dtmax is $0.01$\,days (vertical dotted line). \textit{Bottom} panel: \dtmax over the uncertainty in \tmax (i.e. significance) for the cadence simulations. The average significance is $0.13$ (vertical dotted line), while very few simulations have a significance greater than $3.0$. The average uncertainty in \tmax ($0.19$\,days) is much larger than the weighted average \dtmax.}
    \label{fig:delta_Tmax_cadence}
\end{figure}

In the top panel of Figure~\ref{fig:delta_Tmax_cadence}, we show the difference in \tmax between the cadence simulations and reference sample (\dtmax). For the cadence simulations, only \textit{gr}-bands are used and all those with restframe cadences $<10$\,days are considered. Surprisingly, we can see that despite the inclusion of simulations with slow cadences (i.e. large gaps of up to 10\,days), the distribution of \dtmax has a standard deviation of $\sim0.5$\,days, with a weighted average of $0.01$\,days (weighted by the uncertainty in \tmax of the simulations). By looking at the significance (\dtmax$/\sigma_{T_{\rm max}}$; bottom panel of Figure~\ref{fig:delta_Tmax_cadence}), we see that most are relatively small, with an average of $0.13$, while very few simulations have a significance greater than $3.0$.

\begin{figure}
	\includegraphics[width=\columnwidth]{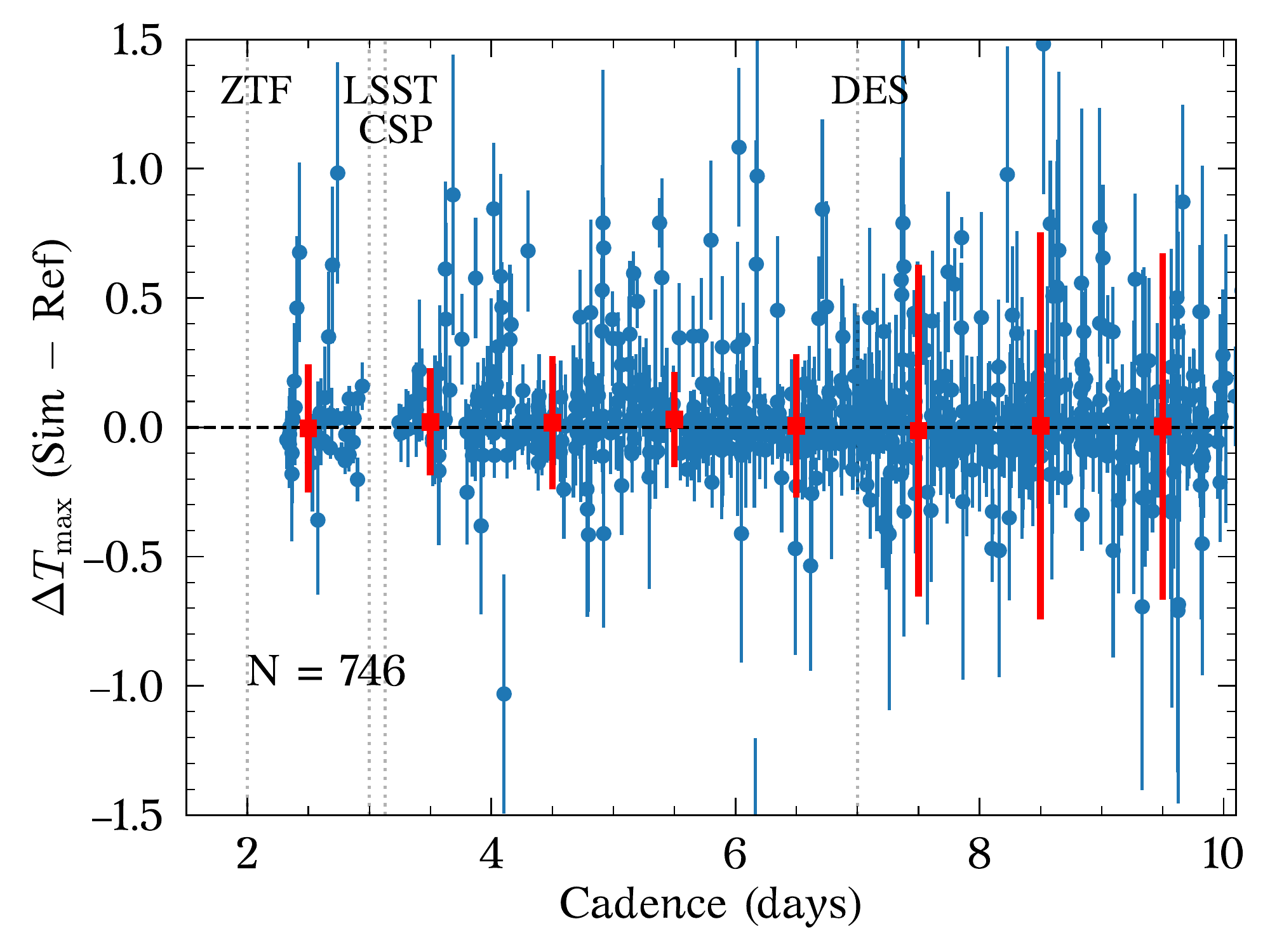}
    \caption{Difference in \tmax between the cadence simulations (Section~\ref{subsec:cadence}) and reference sample (\dtmax) as a function of restframe cadence. The binned average and standard deviation of \dtmax are shown as red squares with error bars. The average observed cadences for ZTF (2\,days), LSST (assuming 3-days baseline cadence per filter), CSP (3.1\,days for the reference sample) and DES (7\,days; \citealt{Brout2019}) are shown as vertical dotted lines. Note that the restframe cadence is faster by a factor of $1+z$ than the observed cadence. N is the total number of simulations.}
    \label{fig:delta_Tmax_vs_cadence}
\end{figure}

This accurate estimation of \tmax can be explained by having small uncertainties and a good coverage across the light curves, i.e. having at least one observation before and after peak, as this helps anchor the light-curve templates. Looking at Figure~\ref{fig:delta_Tmax_vs_cadence}, we see how the weighted average and standard deviation of \dtmax depend on cadence, shown as red squares with error bars. All weighted means are $<0.1$\,days. For cadences $<7$\,days, the standard deviations are all $<0.3$\,days, while for cadences $>7$\,days, they are $>0.5$\,days, but less than $1$\,day. Note that average observer-frame cadences of ZTF (2\,days) and LSST (assuming 3-days baseline cadence in each band) would have a very small effect on the estimation of \tmax, while some additional scatter in \tmax is expected for cadences such as that of DES (7\,days; \citealt{Brout2019}). Also note that the difference between observer-framer and restframe cadences is less than one day at $z<0.1$

\begin{figure}
	\includegraphics[width=\columnwidth]{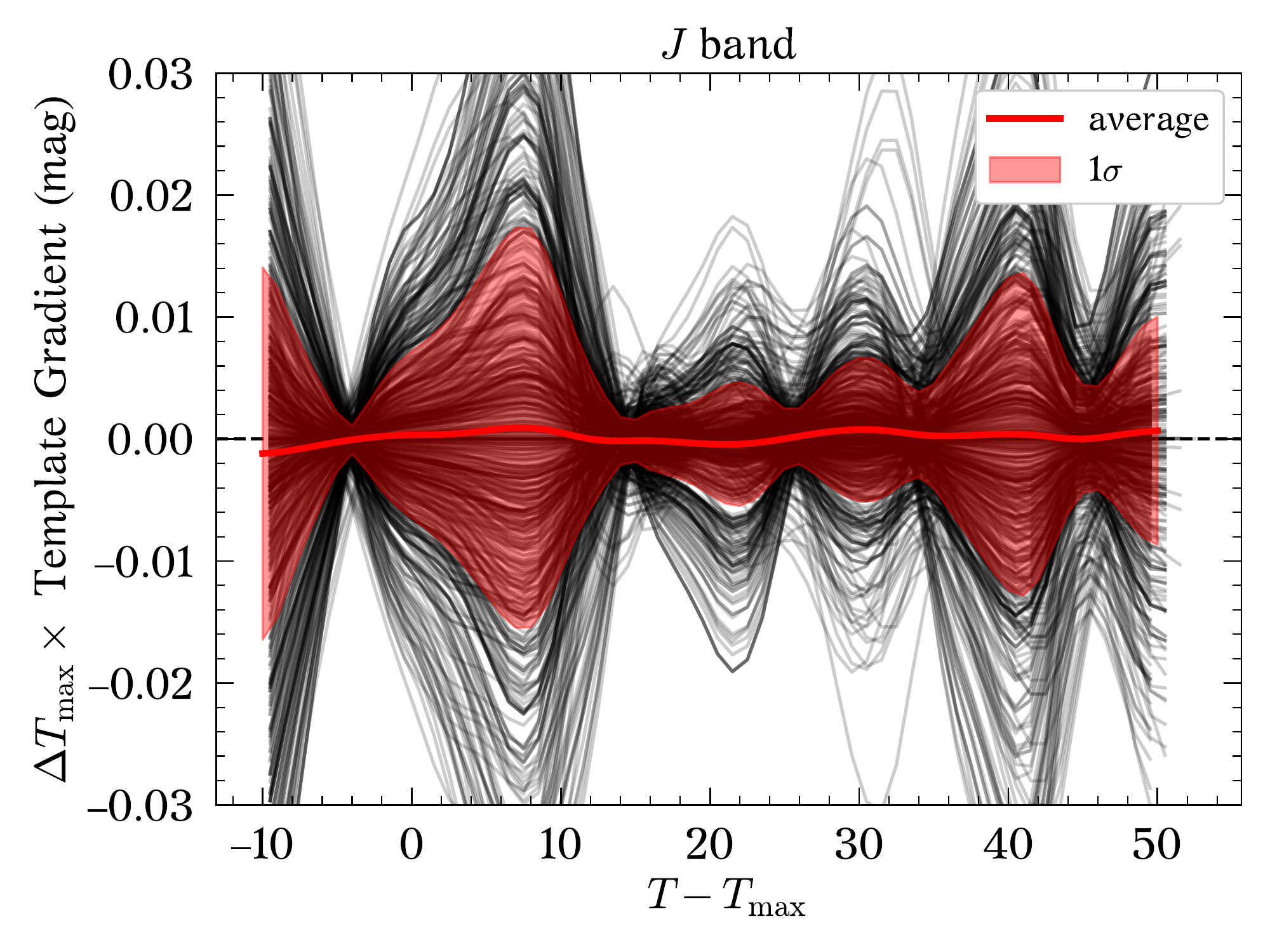}
	\includegraphics[width=\columnwidth]{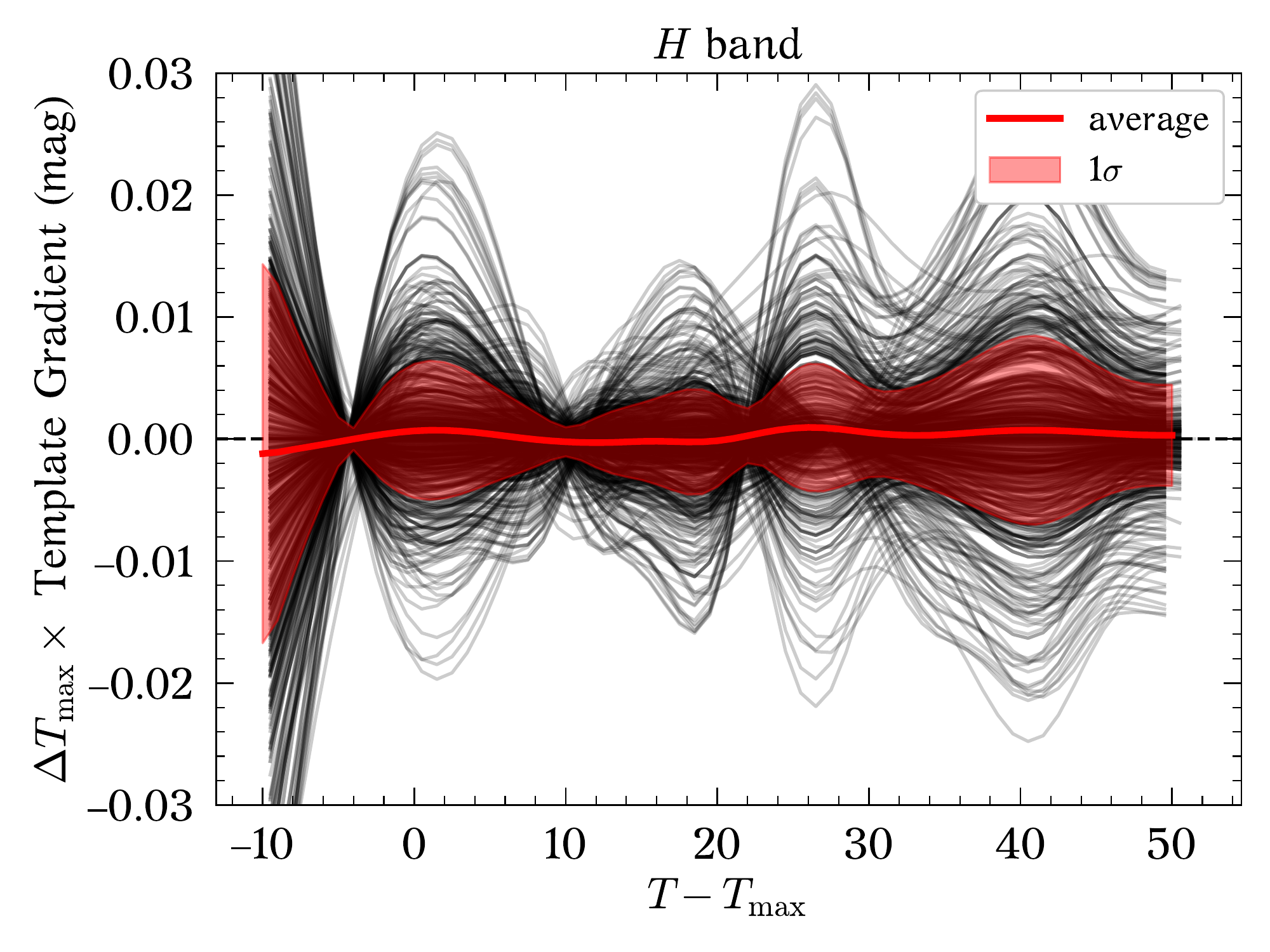}
    \caption{This figure shows the effect that restframe cadence (up to $10$\,days) in the optical \textit{gr}-bands has on the estimation of the NIR peak magnitudes. The grey lines represent \dtmax multiplied by the gradient of the NIR light-curve template bands as a function of phase, for each simulated SN from Section~\ref{subsec:cadence}. Note that the shape of the templates depends on the value of \sbv, which is obtained from the fits of the \textit{gr}-band light curves. The average NIR peak magnitude offset is shown as a red line for \J (\textit{top} panel) and \H (\textit{bottom} panel) bands. The 1$\sigma$ scatter in NIR peak magnitude (red shaded region) is $<0.02$\,mag in \J band and much smaller for \H band ($<0.01$\,mag). }
    \label{fig:Tmax_gradient_cadence}
\end{figure}

We further test how any uncertainty in \tmax could propagate to \Jmax and \Hmax by multiplying \dtmax with the gradient/slope of the NIR light-curve templates (using the definition of error propagation), resulting in additional scatter. Note that the shape of the templates depends on the value of \sbv, which is obtained from the fits of the \textit{gr}-band light curves from the reference sample. The results are shown in Figure~\ref{fig:Tmax_gradient_cadence}. The average offset is negligible and the scatter is $<0.02$\,mag in \J band for NIR observations between $-10$ and $50$\,days. The scatter is much smaller for \H band ($<0.01$\,mag) than for \J band, as the former has a smoother light-curve shape with smaller gradient (see Figure~\ref{fig:sim_example}). Note that these scatters are even lower around \tmax and can be considered as upper limits as cadences of up to $10$\,days are being considered here. Also note that these values are much smaller than the scatter found in Section~\ref{sec:analysis} (i.e. $\sim0.05$\,mag), although have to be added in quadrature.

\begin{figure}
	\includegraphics[width=\columnwidth]{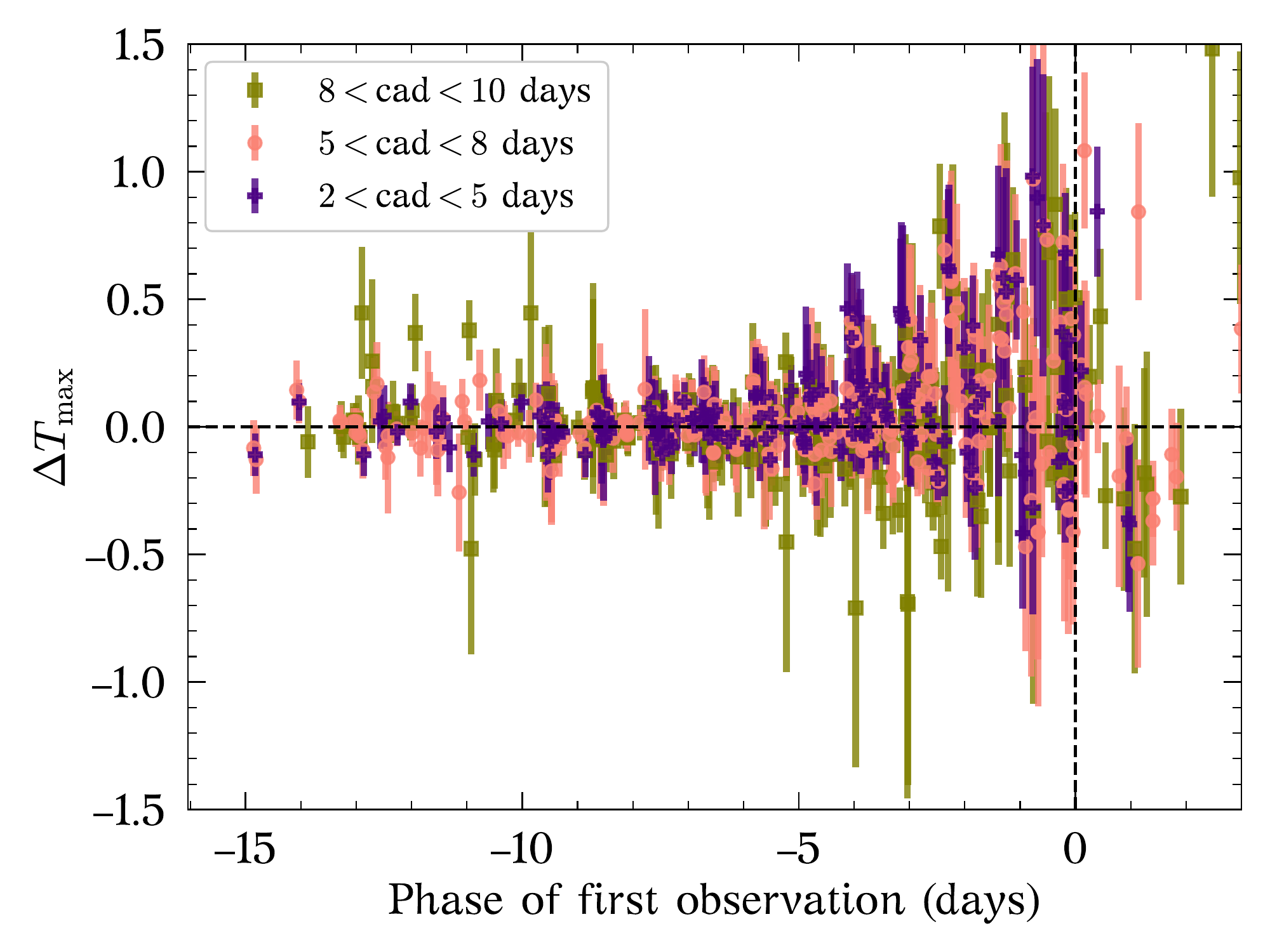}
    \caption{Difference in \tmax between the cadence simulations (Section~\ref{subsec:cadence}) and reference sample (\dtmax) as a function of the phase of the first observation. The simulations are split in three groups according to the cadence (cad) range: $2<\text{cad}<5$\,days (purple crosses), $5<\text{cad}<8$\,days (pink circles) and $8<\text{cad}<10$\,days (gold squares). \tmax is shown as a vertical dashed line. Independent of the cadence, the scatter in \dtmax starts rapidly increasing for first observations with phases $\gtrsim-5$\,days.}
    \label{fig:Tmax_running_cadence}
\end{figure}

Apart from the scatter introduced by different cadences, we can isolate the effect of coverage of the rise of the light curves, which also affects the estimations of \tmax. In Figure~\ref{fig:Tmax_running_cadence}, we show the difference in \tmax between the cadence simulations and reference sample (\dtmax) as a function of the phase of the first observation used for the simulations. One can see that, independent of the cadence of the simulations, the scatter in \dtmax starts rapidly increasing for first observations with phases $\gtrsim-5$\,days. This is really important as it suggests that real observations should ideally cover earlier epochs ($<-5$\,days) in \textit{gr}-bands to provide precise estimations of \tmax. However, observations starting at later epochs can still be used, with the caveat of introducing additional scatter.

\subsection{Signal-to-Noise in the Optical}
\label{subsec:snr}

The effect of S/N is important as it does not only depend on things such as the detector, length of the exposures, etc., but also on the distance and brightness of the objects. To measure how the S/N of optical observations affects the estimation of \tmax, a similar procedure is followed as before. We started by taking \textit{gr}-band light curves of the reference sample and simulated different S/N by multiplying the flux uncertainties in each of the light curves by $x$, for $x = 2, 3, 4, 5, 6, 7$, and randomly sampling new observations using a normal distribution with the new uncertainties. Note that the uncertainties in the CSP observations are relatively small in part due to observing nearby SNe. We use the S/N in \textit{g} band as reference, although the S/N in \textit{r} band is very similar. For each SN, the median S/N is used instead of the mean as the S/N greatly changes as a function of light-curve phase.

\begin{figure}
	\includegraphics[width=\columnwidth]{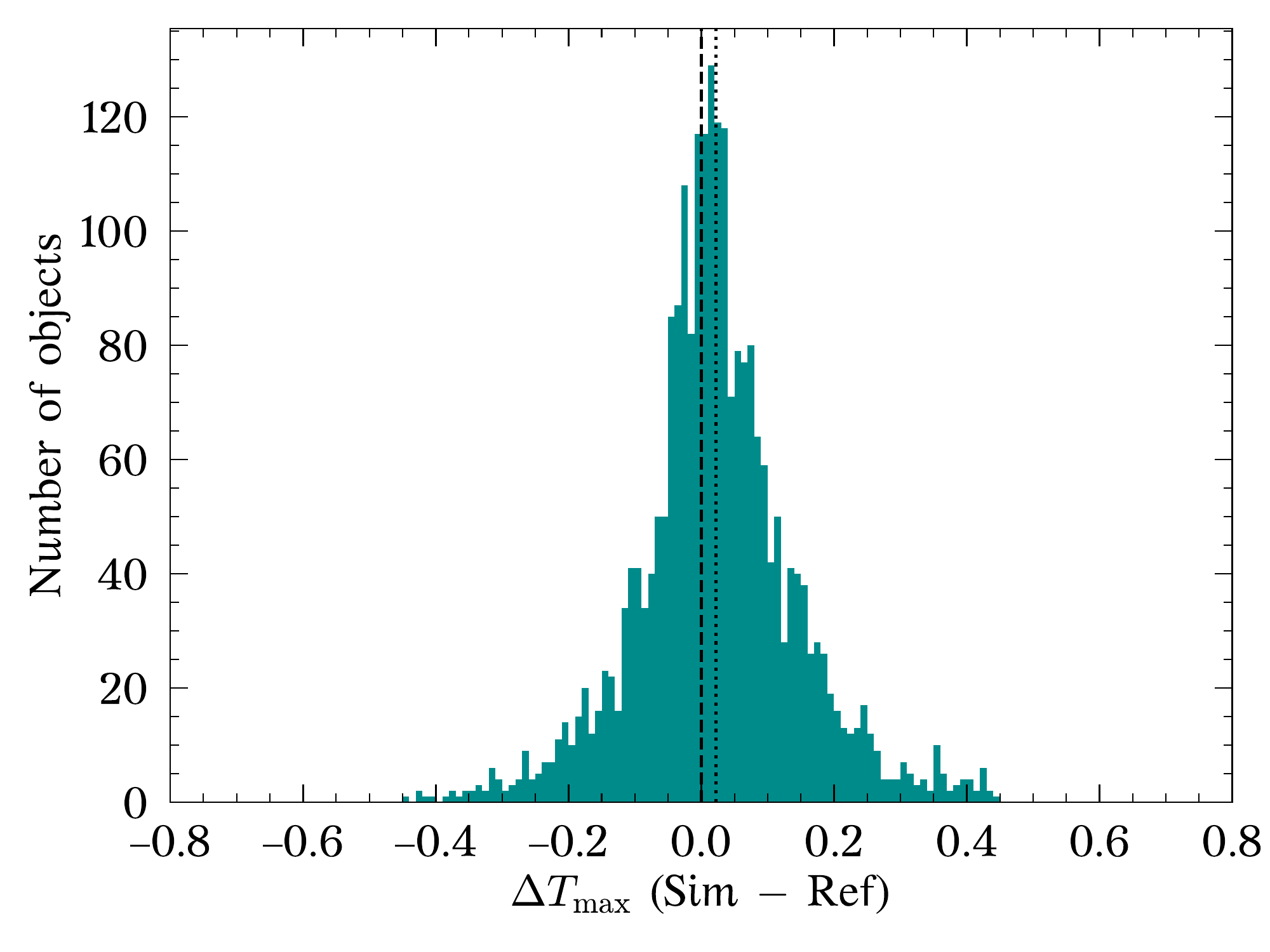}
	\includegraphics[width=\columnwidth]{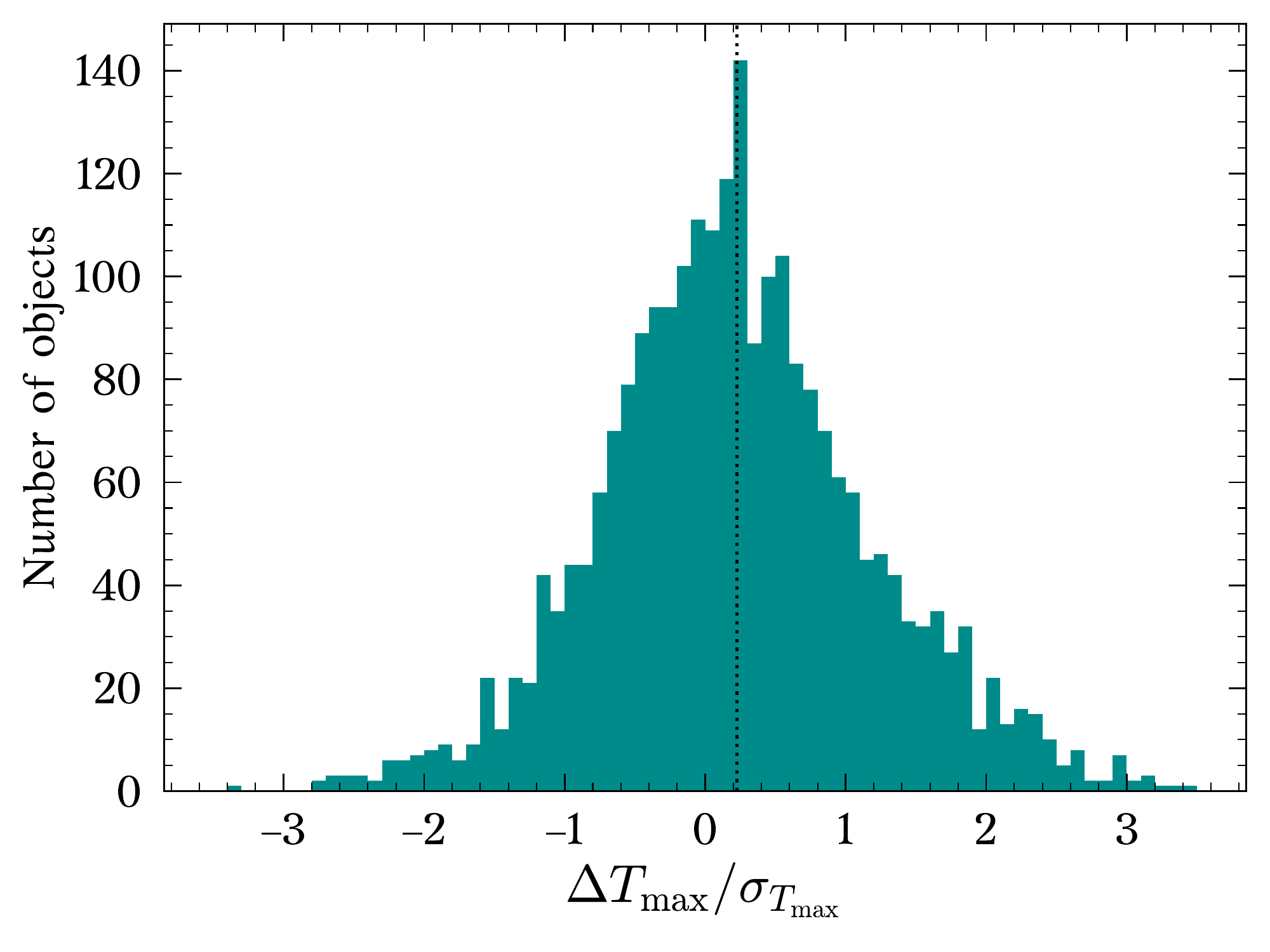}
    \caption{\textit{Top} panel: difference in \tmax between the S/N simulations (Section~\ref{subsec:snr}) and reference sample (\dtmax). The weighted average \dtmax is $-0.007$\,days (vertical dotted line). \textit{Bottom} panel: \dtmax over the uncertainty in \tmax (i.e. significance) for the S/N simulations. The average significance is $0.23$ (vertical dotted line), while very few simulations have a significance greater than $3.0$. The average uncertainty in \tmax ($0.12$\,days) is larger than the weighted average \dtmax.}
    \label{fig:delta_Tmax_snr}
\end{figure}

After calculating the difference in \tmax between the S/N simulations and reference sample, we find relatively similar results compared to those in Section~\ref{subsec:cadence}. The distribution of \dtmax has a standard deviation of $0.13$\,days, with a weighted average of $0.02$\,days (top panel of Figure~\ref{fig:delta_Tmax_snr}), while the average significance in \dtmax is $0.23$ with almost no simulations with a significance greater than $3.0$ (bottom panel of Figure~\ref{fig:delta_Tmax_snr}). 

\begin{figure}
	\includegraphics[width=\columnwidth]{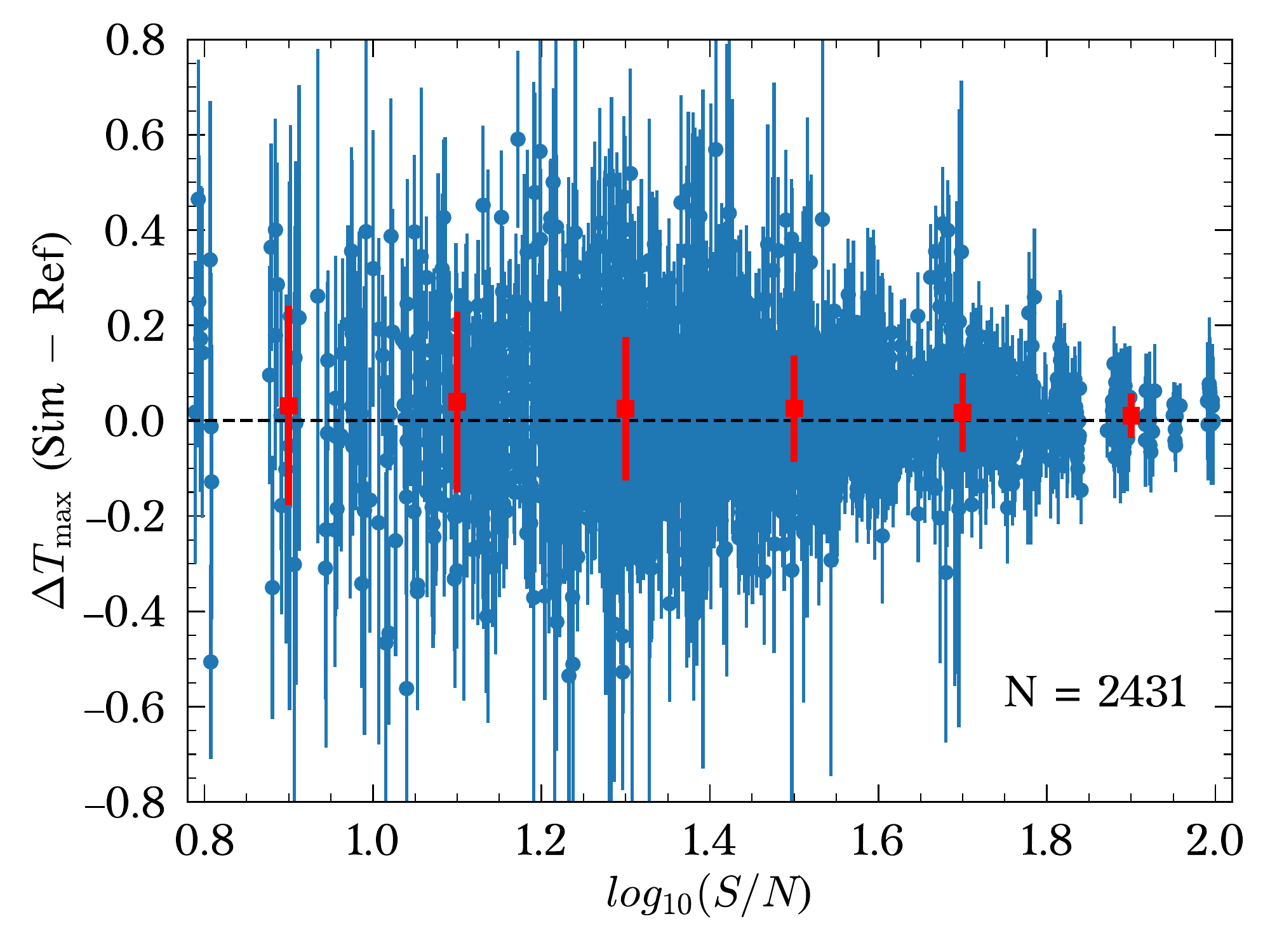}
    \caption{Difference in \tmax between the S/N simulations (Section~\ref{subsec:snr}) and reference sample as a function of S/N. The binned weighted mean and standard deviation of \dtmax are shown as red squares with error bars. N is the total number of simulations.}
    \label{fig:delta_Tmax_vs_snr}
\end{figure}

Figure~\ref{fig:delta_Tmax_vs_snr} shows \dtmax as a function of S/N. This figure shows that the scatter in \dtmax slowly increases as the S/N decreases (note the logarithmic scale in the $x$-axis). Despite the decrease in S/N of the simulations, \snoopy is still able to retrieve a relatively accurate estimation of \tmax. This can be explained by having decent coverage throughout the entire light curves.

\begin{figure}
	\includegraphics[width=\columnwidth]{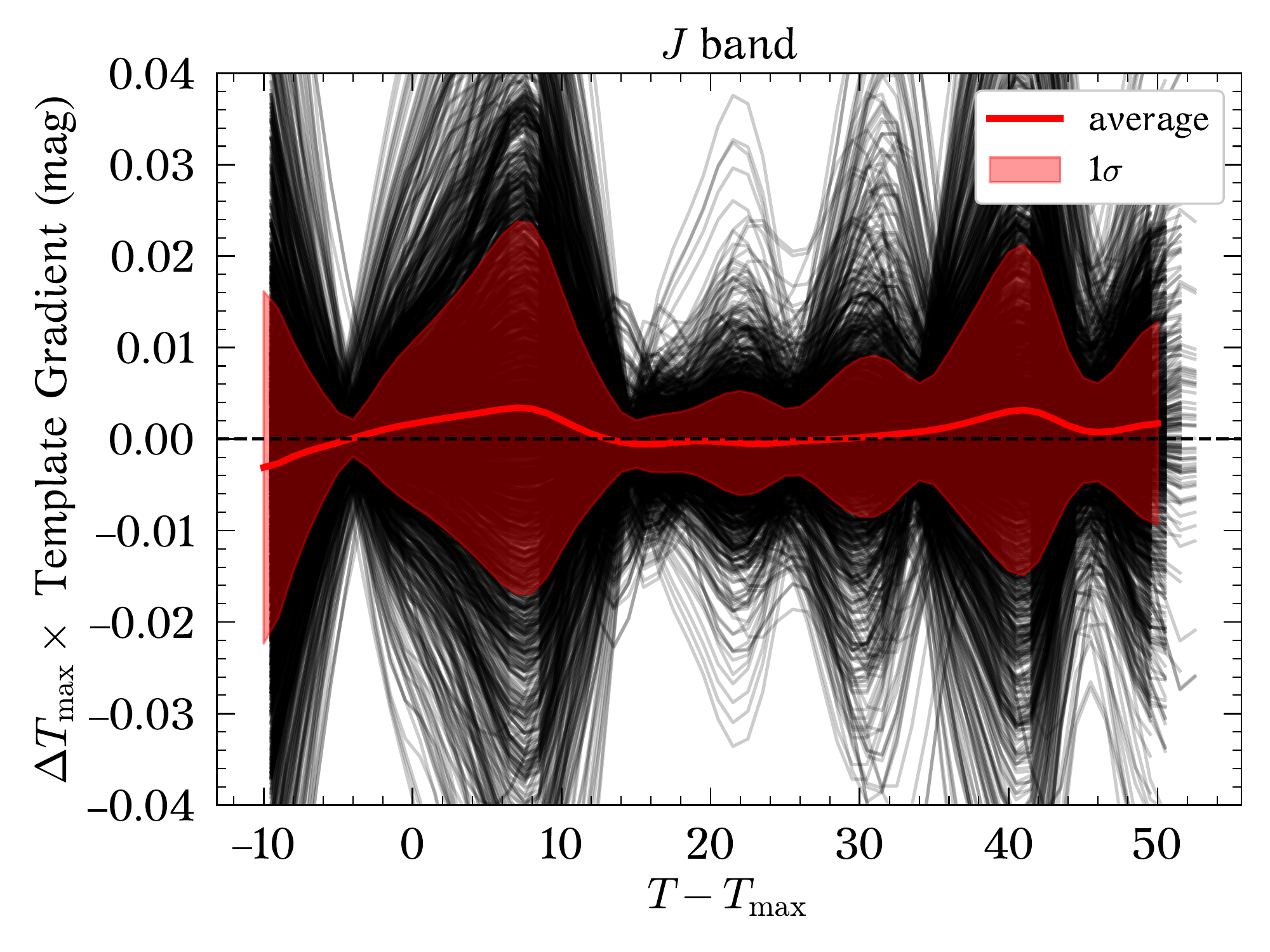}
	\includegraphics[width=\columnwidth]{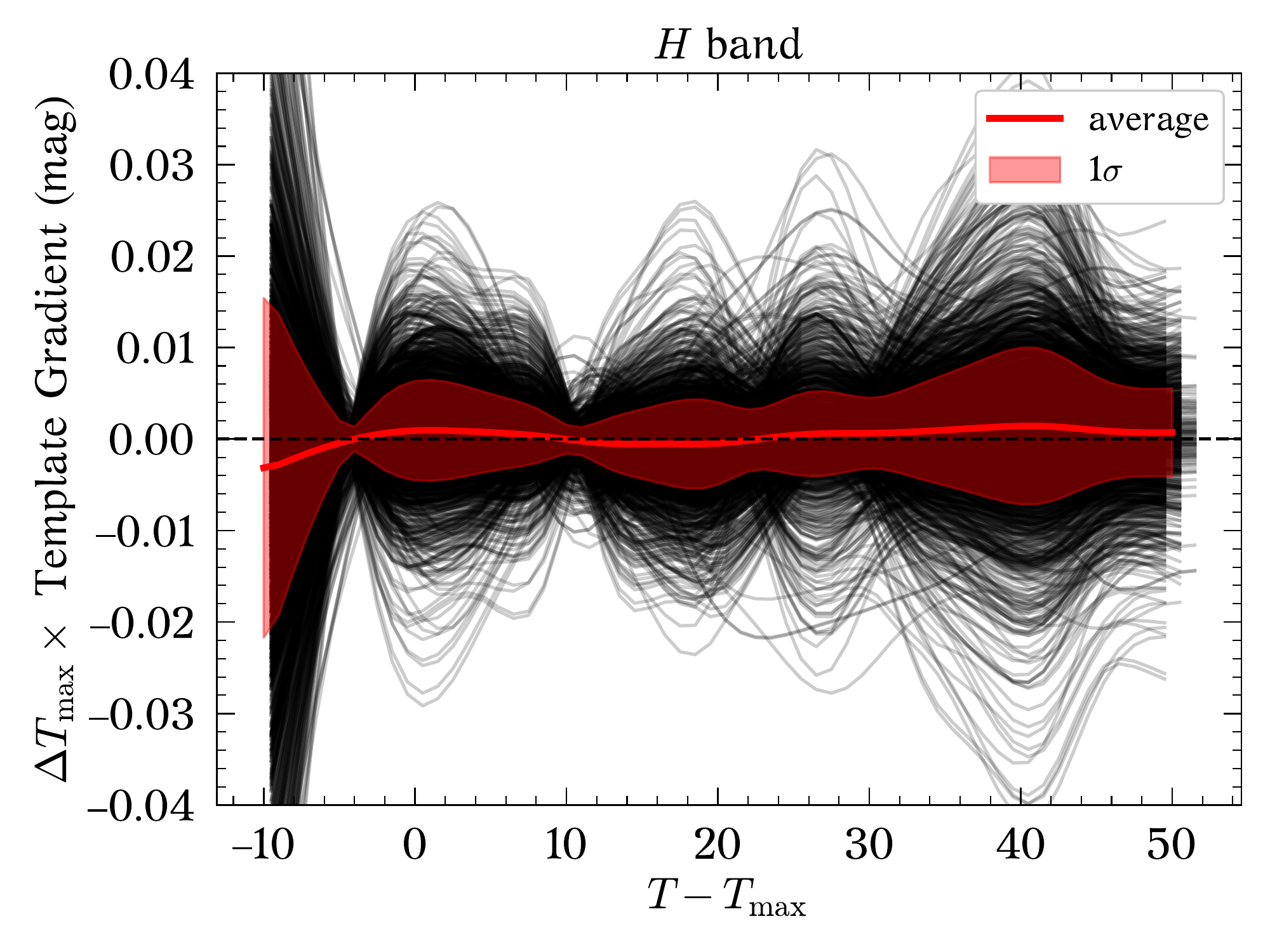}
    \caption{This figure shows the effect that S/N in the optical \textit{gr}-bands has on the estimation of the NIR peak magnitudes. The grey lines represent \dtmax multiplied by the gradient of the NIR light-curve template bands as a function of phase, for each simulated SN from Section~\ref{subsec:snr}. Note that the shape of the templates depends on the value of \sbv, which is obtained from the fits of the \textit{gr}-band light curves. The average NIR peak magnitude offset is shown as a red line for \J (\textit{top} panel) and \H (\textit{bottom} panel) bands. The 1$\sigma$ scatter in NIR peak magnitude (red shaded region) is $\lesssim0.02$\,mag for \J band and $<0.01$\,mag for \H band.}
    \label{fig:Tmax_gradient_snr}
\end{figure}

We also test how the uncertainty in \tmax propagates to \Jmax and \Hmax. The results are shown in Figure~\ref{fig:Tmax_gradient_snr}. Although \Jmax seems to have some slight offset, this is very small ($<0.005$\,mag). The effect of S/N only introduces an additional scatter $\lesssim0.02$\,mag in the \J band and $<0.01$\,mag in the \H band, being even lower around \tmax (smaller than the scatter found in Section~\ref{sec:analysis}). These are similar to the dispersion found in Section~\ref{subsec:cadence}. As we are including simulations with S/N as low as $\sim10$, these dispersions can also be considered as upper limits. Observations of low-$z$ SNe~Ia, such as those observed by the CSP, typically have small photometric uncertainties due to their intrinsically bright nature. 

From this set of tests, we conclude that \snoopy can give an accurate estimation of \tmax, and therefore, \Jmax and \Hmax, for different cadences and S/N in \textit{gr}-bands. Additionally, \H band is less affected by uncertainties in the estimation of \tmax, compared to \J band, due the smooth and relatively flat shape of the SN light curve at these wavelengths. 

\subsection{Signal-to-Noise in the NIR}
\label{subsec:nir_snr}

The effect of S/N is also important for NIR observations of SNe~Ia, especially as the S/N is lower at these wavelengths compared to the optical. To test the effect of S/N of NIR observations in the estimation of the NIR peaks, we proceed in a similar fashion as described in the previous section. However, this time the complete \textit{gr}-band light curves are used together with the closest NIR epoch to \tmax. The uncertainty in the NIR photometric point is then varied to simulate different S/N and the light curves are fitted to estimate \Jmax and \Hmax.

The average S/N of our reference sample is $54$ and $39$ in \J and \H bands, respectively. These are relatively low compared to the average S/N of $\sim100$ in \textit{g} band and SNe~Ia at higher redshift would have even lower values. The CSP sample contains thirteen SNe~Ia with \J-band data at $z>0.08$ with an average S/N of $\sim21$ and only eight objects with \H-band data with an average S/N of $\sim8$. There is not only a large difference in S/N between low-$z$ ($\lesssim0.05$) and high-$z$ SNe ($\gtrsim0.08$), but also between NIR bands.

\begin{figure}
	\includegraphics[width=\columnwidth]{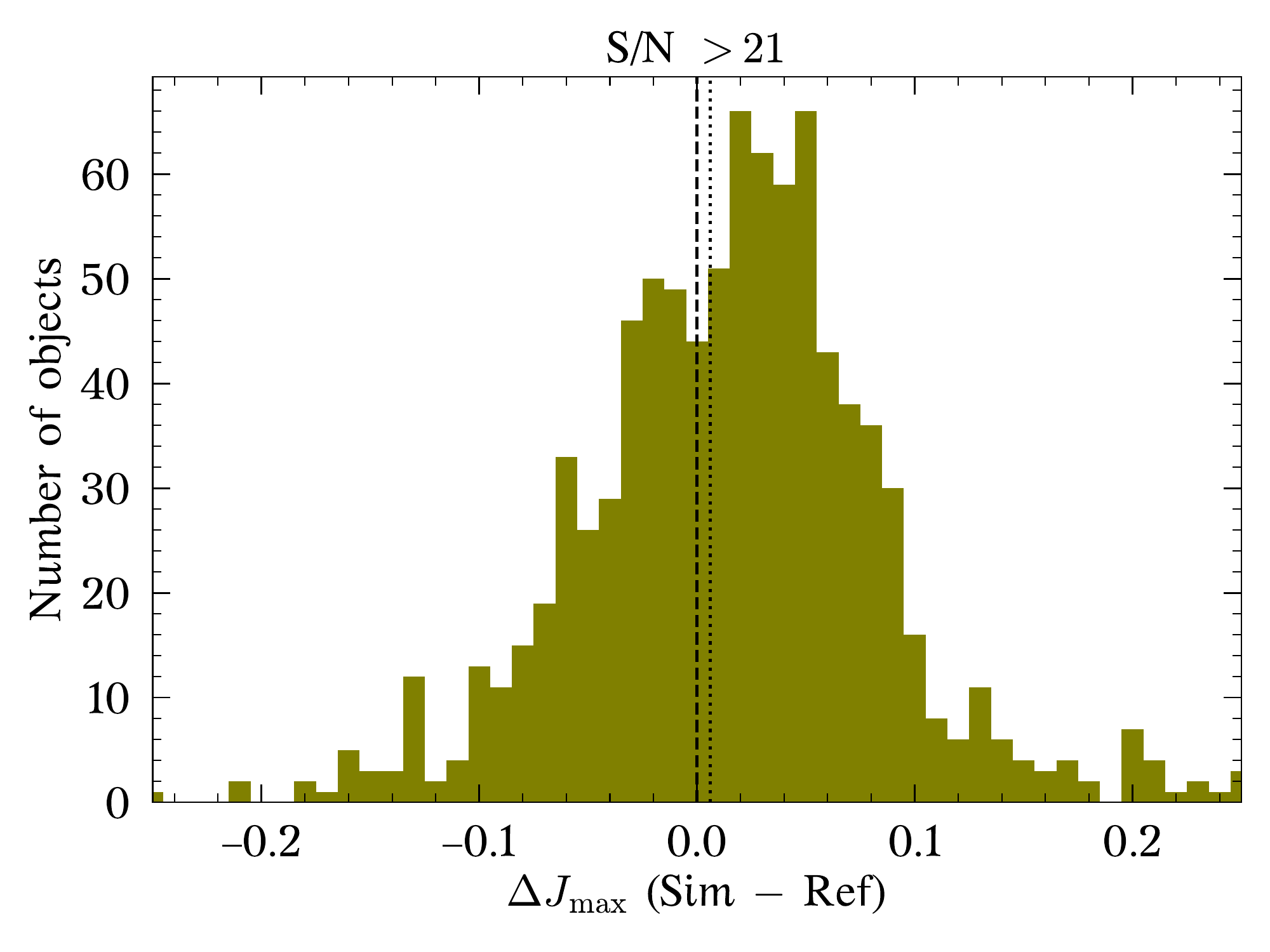}
    \caption{Residual in \Jmax between NIR S/N simulations with \J-band S/N$>21$ (Section~\ref{subsec:nir_snr}) and the reference sample ($\Delta$\Jmax). The weighted mean and standard deviations are $0.006$\,mag and $0.070$\,mag, respectively.}
    \label{fig:delta_Jmax_nir_snr}
\end{figure}

Figure \ref{fig:delta_Jmax_nir_snr} shows the residual in \Jmax between NIR S/N simulations with \J-band S/N$>21$ and the reference sample ($\Delta$\Jmax). The weighted mean and standard deviations of $\Delta$\Jmax are $0.006$\,mag and $0.070$\,mag, respectively. The scatter is larger than the one found in Section~\ref{sec:analysis} ($\sim0.05$\,mag), which was expected. The weighted mean and standard deviations of $\Delta$\Hmax are $0.014$\,mag and $0.107$\,mag, respectively, for simulations with S/N$>8$. These are much larger compared to the \J band and is due to the lower S/N.

\begin{figure}
	\includegraphics[width=\columnwidth]{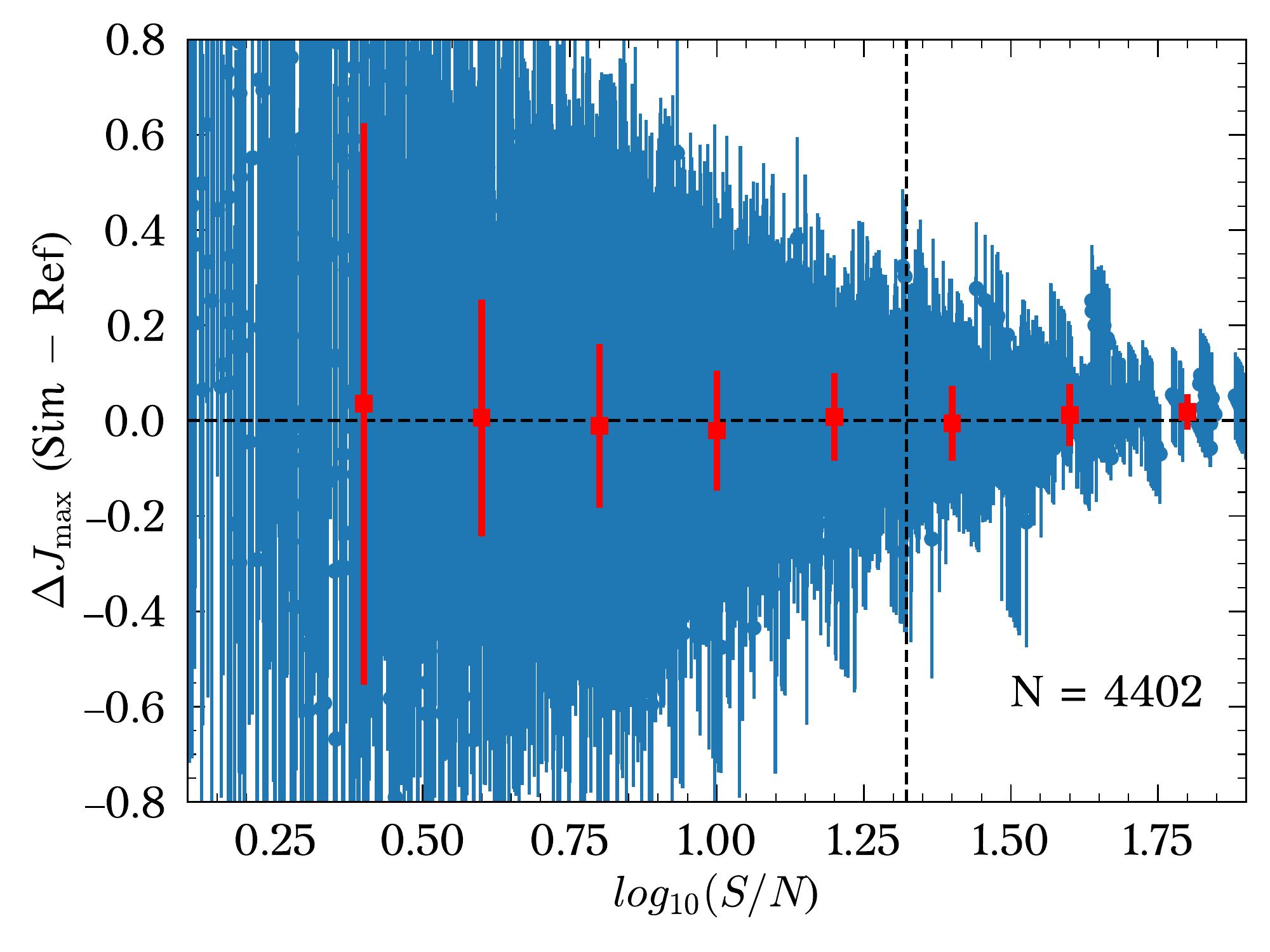}
	\includegraphics[width=\columnwidth]{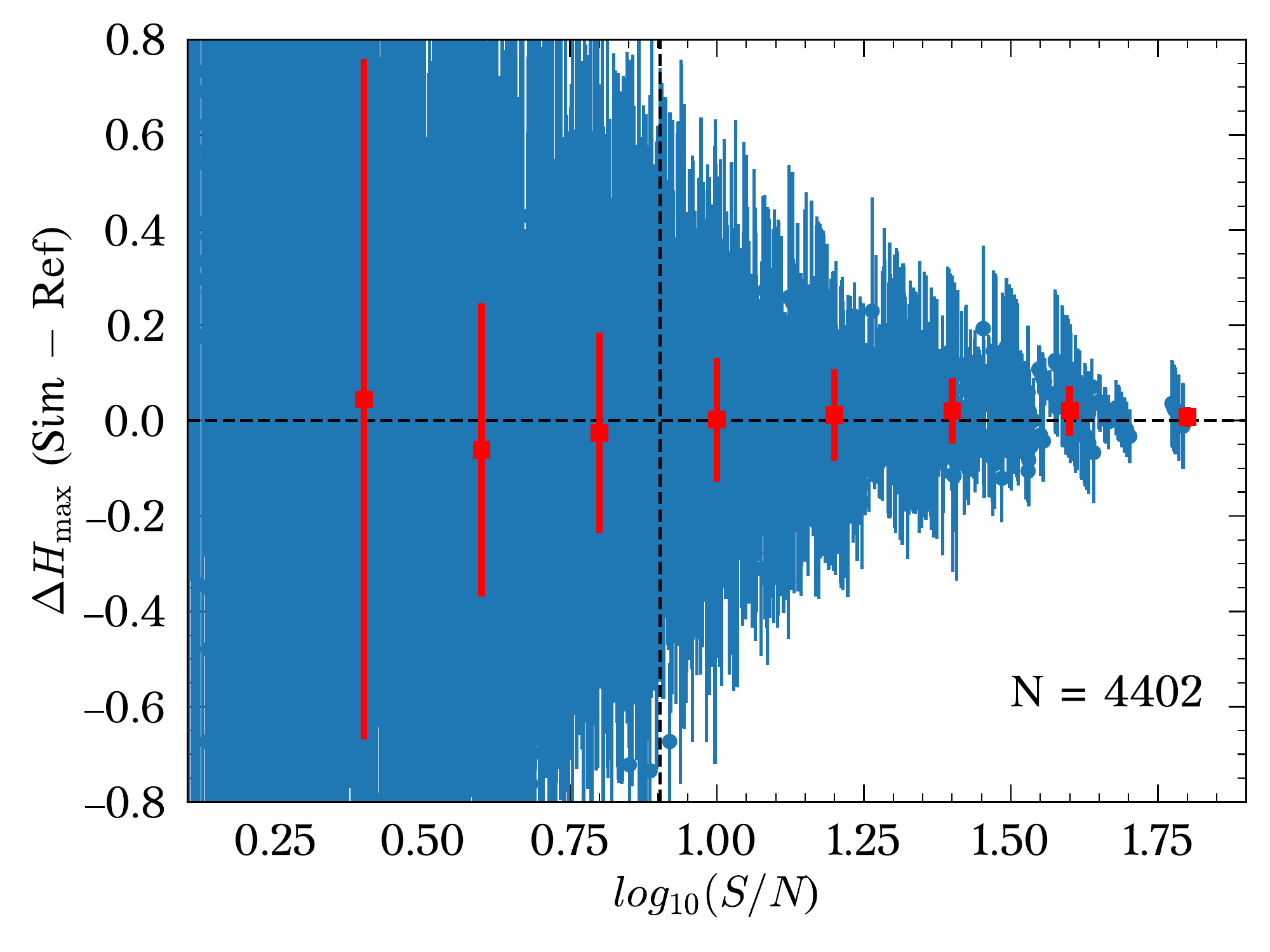}
    \caption{Difference in \Jmax (\textit{top} panel) and \Hmax (\textit{bottom} panel) between the NIR S/N simulations (Section~\ref{subsec:nir_snr}) and reference sample as a function of S/N. The binned median and standard deviations are shown as red squares with error bars. Simulations with all S/N values are shown in here. The average S/N of CSP SNe~Ia at $z>0.08$ are shown as vertical lines (21 and 8 for \J and \H bands, respectively). N is the total number of simulations.}
    \label{fig:delta_NIRmax_vs_snr}
\end{figure}

In Figure~\ref{fig:delta_NIRmax_vs_snr}, we show the $\Delta$\Jmax (\textit{top} panel) and $\Delta$\Hmax (\textit{bottom} panel) as a function of S/N. The scatter in both bands is similar for the same S/N, although the average S/N is lower in \H band. This test shows the importance of having high S/N observations in the NIR. If the aim is having accurate estimation of \Jmax and \Hmax for SNe~Ia at $z\sim0.1$, NIR observations with better S/N than those obtained by CSP must be achieved (see Figure~\ref{fig:delta_NIRmax_vs_snr}). Note that, however, several of the CSP observations were obtained with a $1$\,meter telescope (Swope), so using $2$\,meter-class telescopes or better is ideal. Also note that, in the restframe, the \J-band light curves of SNe~Ia are intrinsically brighter than their \H-band light curves, making \J-band observations better suited for measuring distances at high $z$.

\section{Near-Infrared Distances}
\label{sec:nir_distances}

\begin{figure}
	\includegraphics[width=\columnwidth]{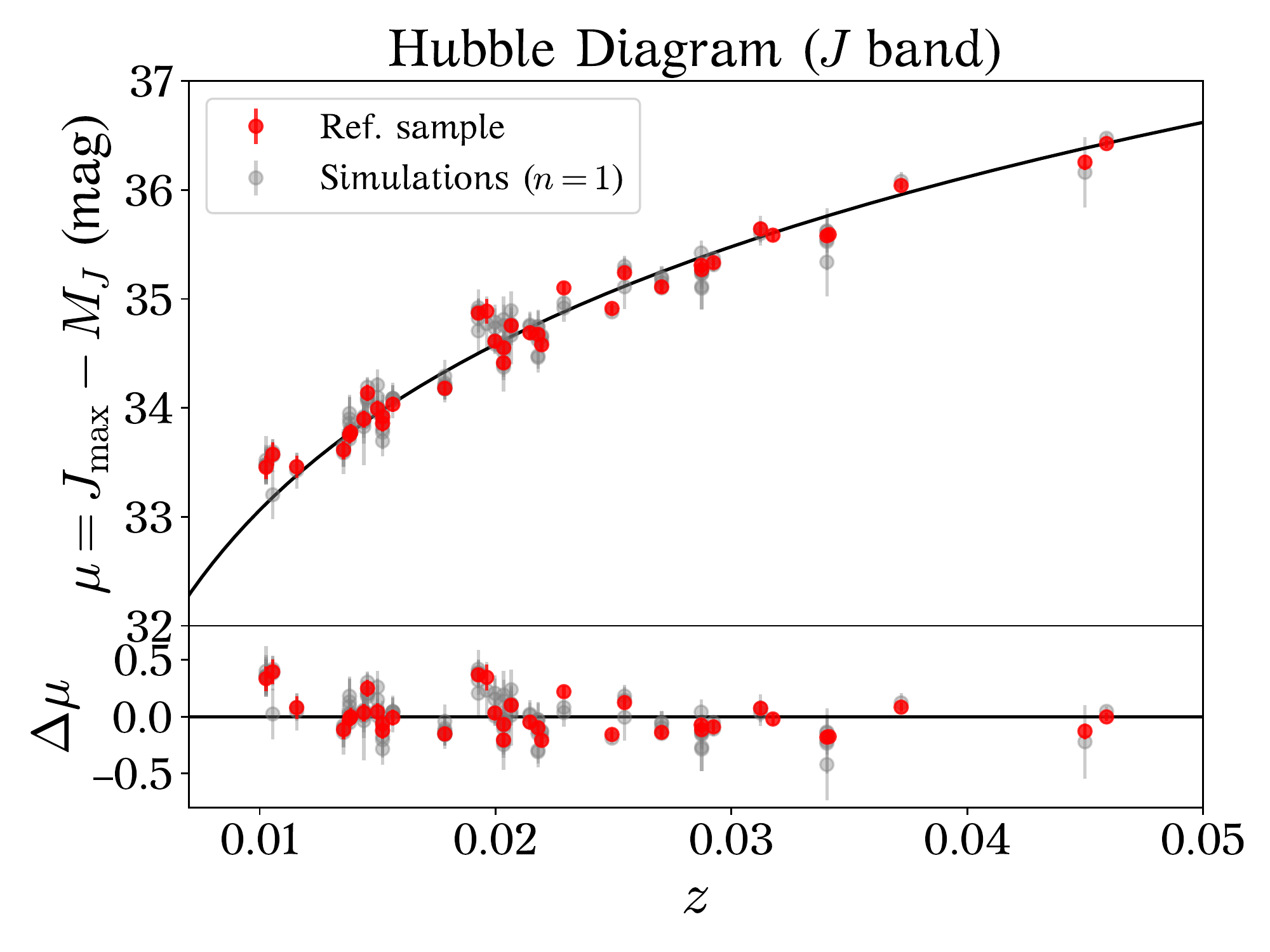}
    \caption{Hubble diagram in \J band. Only SNe~Ia with $z>0.01$ were used as the contribution from peculiar velocities is relatively small at these redshifts. The reference sample was used to fit $H_0$ and the intrinsic dispersion ($\sigma_{\rm int}$), keeping the peak absolute magnitude in $J$ band ($M_J=-18.5$\,mag) fixed. The RMS for the reference sample (red) and the simulations with $n = 1$ (grey) are $0.166$\,mag and $0.180$\,mag, respectively.}
    \label{fig:hubble_diagram}
\end{figure}

The final step in this work is to calculate the precision in the distance estimations from the simulations of Section~\ref{sec:analysis}. Assuming that SNe~Ia are standard candles in the NIR, the peak apparent magnitude is the only parameter necessary to calculate distances and its uncertainty is directly propagated to the measured distance:

\begin{equation}
    \mu = m_{\rm max} - M,
\end{equation}

\noindent where $\mu$ is the distance modulus, $m_{\rm max}$ is the peak apparent magnitude in a NIR band (e.g., \Jmax or \Hmax) and $M$ is the peak absolute magnitude in that same band. To calculate distances, a cosmological model needs to be fitted. For simplicity, we assume a flat $\Lambda$CDM cosmology and fix the value of $M$ ($M_J = M_H = -18.5$\,mag), fitting only $H_0$ and the intrinsic dispersion of SNe~Ia ($\sigma_{\rm int}$). Only SNe at $z>0.01$ are used as the contribution from peculiar velocities is relatively small at these redshifts. This reduces our reference sample to 36 objects. The resulting Hubble diagram in \J band is shown in Figure~\ref{fig:hubble_diagram} (red circles). The Hubble residuals have an RMS of $0.166$\,mag, while $\sigma_{\rm int}=0.14$\,mag was obtained.

To calculated distances with the simulations, we use the values of $H_0$ and $\sigma_{\rm int}$ obtained with the reference sample, and apply the offsets found in Section~\ref{sec:analysis}:

\begin{equation}
    \mu = [m_{\rm max} + \Delta(b, n, p)] - M,
\end{equation}

\noindent where the offset $\Delta(b, n, p)$ depends on the NIR band ($b$), the number of epochs ($n$) and the phase ($p$) of the epoch closest to \tmax. The uncertainty associated to this offset is added in quadrature. If one assumes that SNe Ia are standard candles in the NIR, one would expect to measure the same offsets, as found in Section~\ref{sec:analysis}, in \Jmax and \Hmax for a different sample of SNe Ia. Therefore, these can be used as a correction term.

The \J-band Hubble diagram for simulations with $n=1$ is shown in Figure~\ref{fig:hubble_diagram} (grey circles). Using the simulations with phase between $-5$\,days and $15$\,days, where a low scatter in \Jmax was found (see Section~\ref{sec:analysis}), an RMS of $0.180$\,mag is obtained. Although the simulations have larger scatter than the reference sample, the difference is relatively small ($0.014$\,mag).

In the case of the \H band, the reference sample and the simulations have Hubble residuals RMS of $0.149$\,mag and $0.147$\,mag, respectively. The RMS values are very similar, being slightly smaller for the simulations (a negligible difference of $0.002$\,mag). Although a larger scatter was expected for the simulations, note that sample used to build the Hubble diagram is not exactly the same as that used in Section~\ref{sec:analysis}. If the offsets found in Section~\ref{sec:analysis} are not applied, very similar results are found, only increasing the RMS in $\sim0.002$\,mag for \J and \H bands. Therefore, we believe that these offsets might not be necessary to apply.

As expected, the scatter in \H band is lower than in the \J band. In addition, for the \H band $\sigma_{\rm int}=0.12$\,mag, smaller than for the \J band. If the simulations are used to fit $H_0$ and $\sigma_{\rm int}$ instead of the reference sample, we find the same RMS values. The values of $H_0$ and $\sigma_{\rm int}$ do change, but the differences are insignificant ($\lesssim1.5\sigma$) for both \J and \H bands.

By using simulations with 2 and 3 NIR epochs, the scatter is reduced to $0.178$\,mag and $0.168$\,mag for the \J band and $0.138$\,mag and $0.127$\,mag for the \H band, respectively. Although the reduced scatter is expected, we did not expect to have lower scatter for the simulation than for the reference sample in the \H band. This could be caused by limitations in our analysis. For instance, \snoopy was trained on a sub-sample of the SNe Ia from CSP, which can possibly bias the estimation of light-curve parameters, artificially producing a better RMS in the simulations. This will be studied in the future as other NIR light-curve fitters and/or accurate theoretical light-curve models of SNe Ia become available.

The Hubble residuals RMS values found in this work are smaller than those found by \citet{Uddin2020}, which have values of $0.183$\,mag in both \J and \H bands, for SNe~Ia at $z>0.01$. However, they did not include \cspii SNe and the cuts applied to their sample are different to those applied in this work (Section~\ref{subsec:sample_cuts}). Nonetheless, we have shown here that sparse NIR observations of SNe~Ia can be used to measure accurate distance and obtained comparable scatter to those found using well sampled NIR light curves.

\section{Summary \& Conclusions}
\label{sec:conclusions}

In this work, we have explored whether it is possible to obtain accurate \J- and \H-band peak magnitudes from just a few NIR epochs plus good \textit{gr}-band coverage. To this end, we used the CSP SN~Ia sample, the most comprehensive samples with extensive optical to NIR (\textit{uBgVriYJH}) coverage that exist to date. Combining \cspi and \cspii, we gathered a total of 336 SNe~Ia, to which we applied a set of quality cuts, reducing the number to 50 SNe, comprising our reference sample. The objects were fitted with \snoopy, using the \maxmodel model, to produce a set of reference values for \Jmax and \Hmax, using all available bands.

A set of simulations was created by selecting $n$ (where $n = 1, 2, 3$) coeval epochs in \J and \H bands (the reddest available bands), using combinations without repetition, and using all the available photometry in the \textit{g} and \textit{r} bands. The resulting simulated \textit{grJH}-band light-curves were then fitted with \snoopy, with the same configuration used for the reference sample, to obtain estimations of \Jmax and \Hmax. We then proceeded to compare the NIR peak magnitudes between the simulations and reference sample, finding relatively good agreement in general (residuals $<0.01$\,mag). For simulations with $n = 1$, we found that NIR epochs between $-5$ to $15$ days with respect to \tmax only introduce an additional scatter of $\sim0.05$\,mag in both \Jmax and \Hmax, with respect to our reference sample. Similar results where found when using \J and \H bands independently, i.e. \textit{grJ} and \textit{grH}.

For simulations with $n = 2, 3$, we find that the scatter in the estimation of \Jmax and \Hmax is reduced and that the most relevant factor when estimating the NIR peak magnitudes is the time of the closest NIR epoch to \tmax (metric (i)).

A set of tests were performed to estimate the effect of cadence of the optical light curves in the estimation of \tmax and its uncertainty propagation to \Jmax and \Hmax. The results show that cadences up to $10$\,days only introduce an additional scatter of $<0.02$\,mag in \Jmax, being much smaller in \Hmax ($<0.01$\,mag). The effect of optical S/N in the estimation of \Jmax and \Hmax was also tested, finding similar scatters. These scatters are smaller than those found in Section~\ref{sec:analysis} and can be considered as upper limits if NIR around \tmax is used (phases between $-5$ to $15$ days). In addition, we tested the effect of S/N in the NIR light curves. For simulations with S/N similar or better than that of CSP SNe~Ia at $z>0.08$, we found larger scatter compared to the analysis of Section~\ref{sec:analysis} ($>0.05$\,mag). From these tests, we conclude that \snoopy is expected to retrieve accurate estimations of NIR peak magnitudes for SNe~Ia out to $z\sim0.1$ provided of NIR observations with better S/N than those obtained by CSP.

This work presents some limitations that need to be considered. For example, \snoopy was trained on a sub-sample of the SNe Ia from CSP, which can possibly bias the estimation of light-curve parameters. The use of theoretical SN Ia models to produce synthetic observations in the NIR would help in this case, although none exist to date. The sample of SNe Ia used (after cuts) is relatively small, limiting the statistics. Only objects with $z\lesssim0.05$ where used, making the extrapolation of this analysis to higher redshifts not completely reliable. Nonetheless, we have shown how standard are SNe Ia in the NIR. Furthermore, this work can be repeated in the future when more SNe with well covered NIR light curves become available (e.g. with the Roman Space Telescope).

The results from this work can be used by the community to develop the strategy of future surveys of SNe Ia, such as that from the Roman Space Telescope. In addition, they provide confidence for our Aarhus-Barcelona FLOWS\footnote{\url{https://flows.phys.au.dk/}} project that aims in using SNe~Ia with public ZTF optical light curves and minimal (one to three) NIR epochs to map out the peculiar velocity field of the local Universe. This will allow us to determine the distribution of dark matter in our home supercluster and test the standard cosmological model by measuring the growth of structure ($fD$) and the local value of $H_0$.


\begin{acknowledgements}
TEMB and LG acknowledge financial support from the Spanish Ministerio de Ciencia e Innovaci\'on (MCIN), the Agencia Estatal de Investigaci\'on (AEI) 10.13039/501100011033 under the PID2020-115253GA-I00 HOSTFLOWS project, and from Centro Superior de Investigaciones Cient\'ificas (CSIC) under the PIE project 20215AT016, and the I-LINK 2021 LINKA20409. 
TEMB and LG are also partially supported by the program Unidad de Excelencia Mar\'ia de Maeztu CEX2020-001058-M.

LG also acknowledges MCIN, AEI and the European Social Fund (ESF) "Investing in your future" under the 2019 Ram\'on y Cajal program RYC2019-027683-I.

The Aarhus SN group is support by a Villum Experiment grant (number 28021) from VILLUM FONDEN and a Project 1 grant (8021-00170B) from the Independent Research Fund Denmark. 

PH acknowledge support by the National Science Foundation (NSF) grant AST-1715133.

The work of the Carnegie Supernova Project has been supported by the NSF under the grants AST0306969, AST0607438, AST1008343, AST1613426, AST1613472 and AST613455.
\end{acknowledgements}

\textit{Software}: 
\texttt{matplotlib} \citep{matplotlib}, 
\texttt{seaborn} \citep{seaborn},
\texttt{numpy} \citep{numpy},
\texttt{pandas} \citep{pandas},
\texttt{scipy} \citep{scipy},
\texttt{emcee} \citep{emcee}, 
\texttt{coner} \citep{corner}, 
\texttt{george} \citep{Ambikasaran2016}, 
\texttt{astropy} \citep{astropy, astropy2},
\texttt{peakutils} \citep{peakutils}.

\bibliographystyle{aa}
\bibliography{flows}

\appendix
\section{Light-Curve Fitter}
\label{app:snoopy}

\snoopy is a versatile light-curve fitter that works on optical and NIR bands. It incorporates different fitting models, such as \maxmodel and \EBVmodel, adapting to the needs of the science. The main difference between both is that the former fits each band independently (using a common \tmax and \sbv for the light-curve templates), while the latter fits a dust extinction law for the host galaxy\footnote{\url{https://users.obs.carnegiescience.edu/cburns/SNooPyDocs/html/models.html}}. Hence, the different models might not necessarily produce the same results. In this Appendix, the two models mentioned above are tested.

\begin{figure}
	\includegraphics[width=\columnwidth]{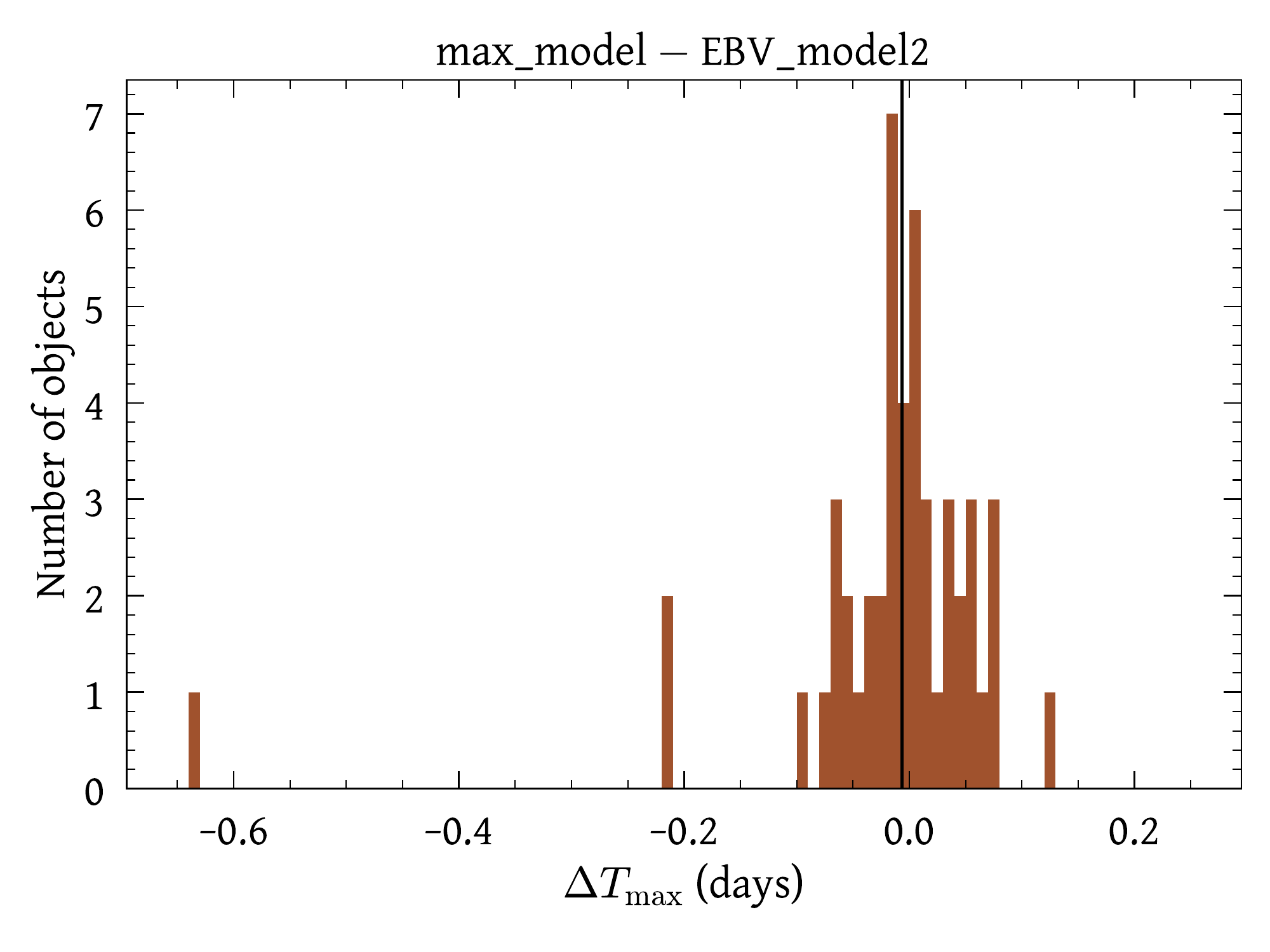}
    \caption{\tmax residuals (\dtmax) between \maxmodel and \EBVmodel, using the reference sample from Section~\ref{subsec:ref_sample}. The weighted mean of \dtmax is $\sim-0.001$\,mag (vertical black line).}
    \label{fig:model_comparison_Tmax}
\end{figure}

The reference sample from Section~\ref{subsec:ref_sample} was fitted with \snoopy using the \maxmodel and \EBVmodel models. From these fits, we see that the bulk of the objects have a difference in \tmax $<0.1$\,days, and there is no significant difference in the average estimation  of \tmax (see Figure~\ref{fig:model_comparison_Tmax}).

\begin{figure}
	\includegraphics[width=\columnwidth]{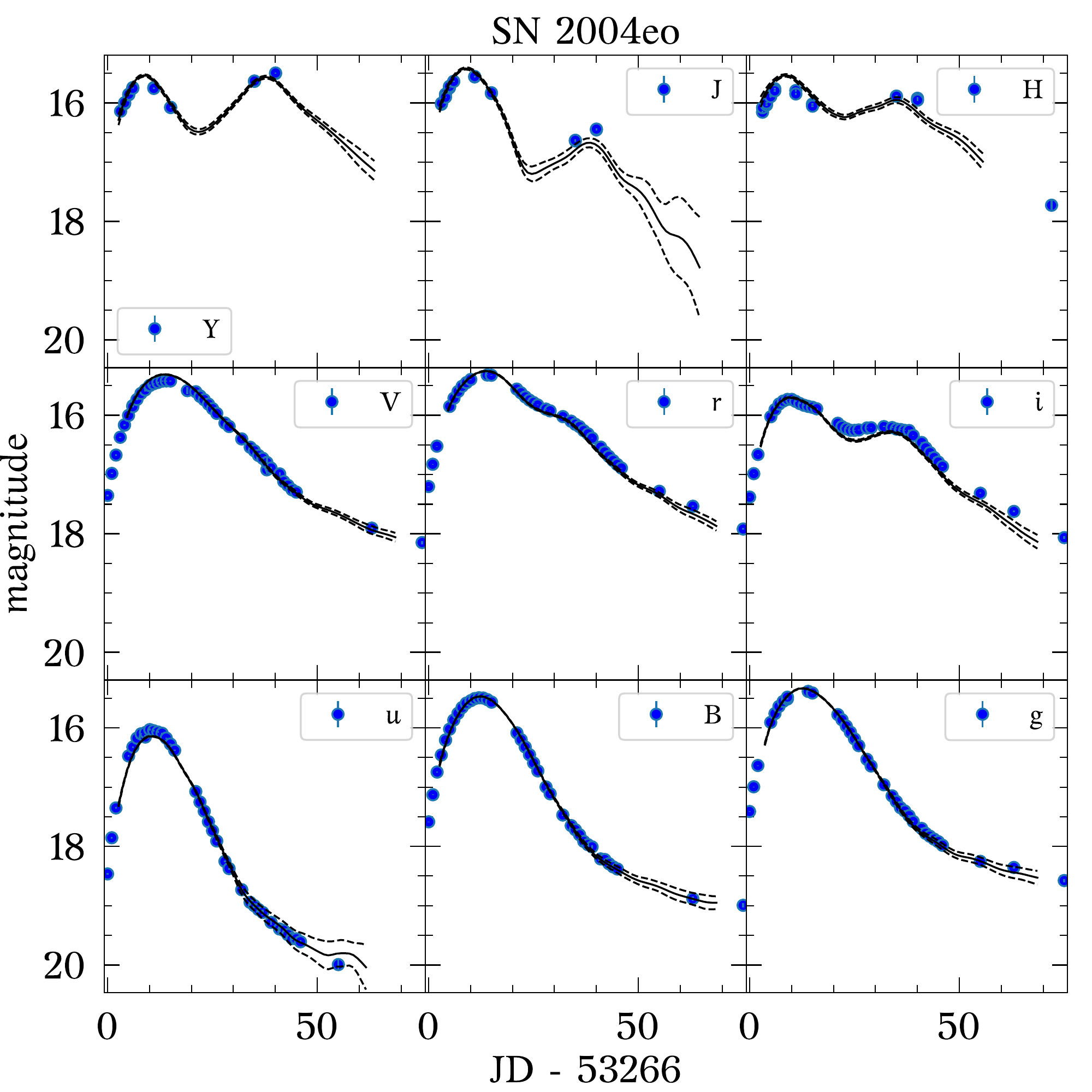}
    \caption{\snoopy fits of SN~2004eo with the \EBVmodel. Note that the \EBVmodel model does not produce reliable fits of the bluest (\textit{u}) and reddest (\textit{H}) filters.}
    \label{fig:fits_comparison}
\end{figure}

\begin{figure}
	\includegraphics[width=\columnwidth]{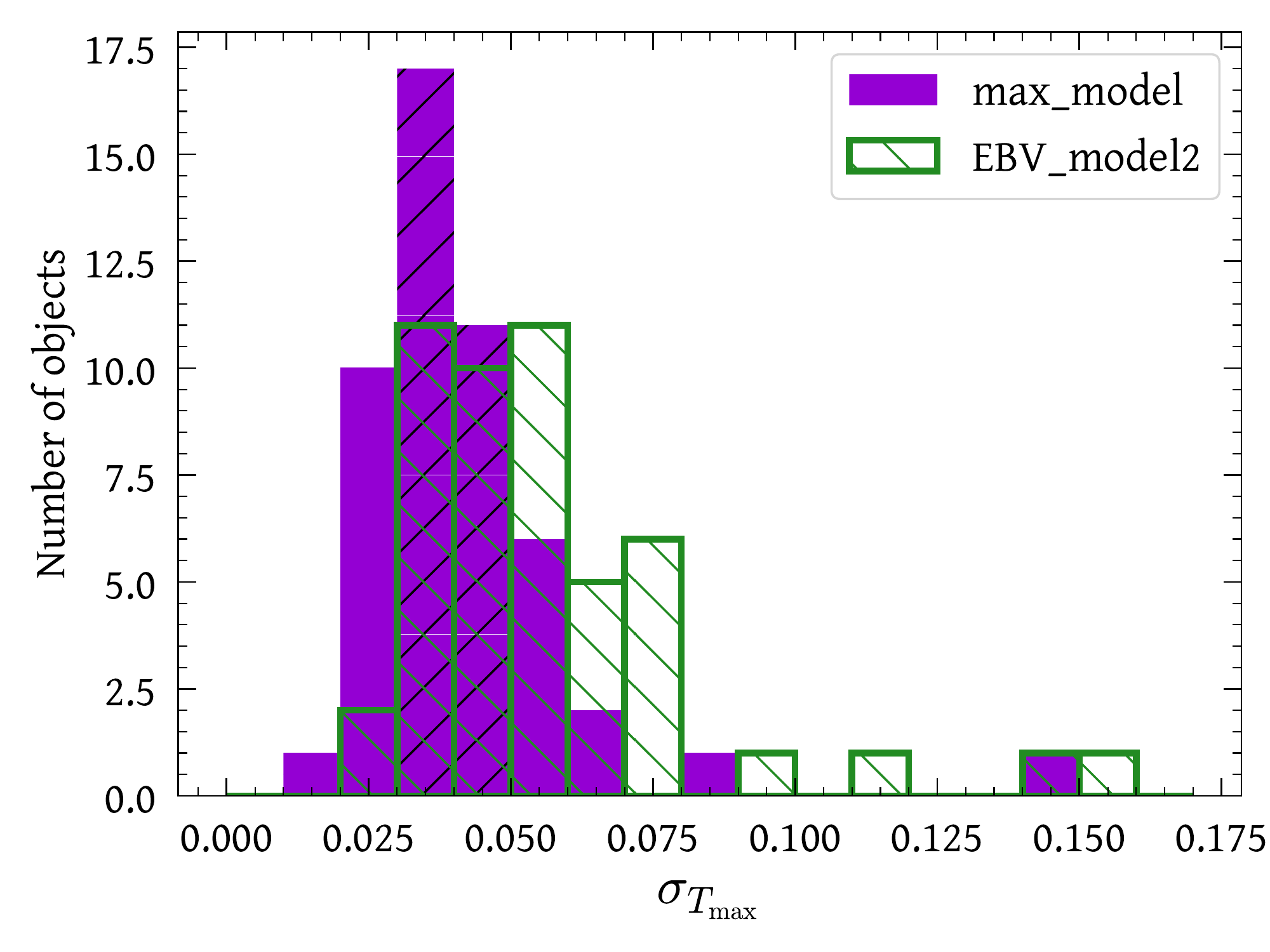}
    \caption{Comparison in the uncertainty of \tmax ($\sigma$) between \maxmodel and \EBVmodel, using the reference sample from Section~\ref{subsec:ref_sample}. Note that the uncertainties from the latter are larger.}
    \label{fig:model_comparison_Tmax_err}
\end{figure}

The object with the largest difference in \tmax is SN~2006X. This object is known to be a highly reddened SN \citep[see, e.g.,][]{Wang2008, Folatelli2010}, which might causes this inconsistency. However, looking at the fits of this and other SNe~Ia, we see a difference between the \snoopy models. An example of the resulting fits for SN~2004eo, using the \EBVmodel model, is presented in Figure~\ref{fig:fits_comparison}. From the fits, one can see that the \EBVmodel model does not produce reliable fits of the bluest (\textit{u}) and reddest (\textit{H}) filters (compared to the fits from Figure~\ref{fig:snoopy_fit}). A similar behaviour is observed in other objects, including SN~2006X. In addition, the \EBVmodel model produces larger uncertainties in the estimation of \tmax than the \maxmodel model (see Figure~\ref{fig:model_comparison_Tmax_err}). Thus, from these tests, we choose to use the \maxmodel model throughout this work.

\section{NIR offsets with single \J and \H bands}
\label{app:extras}

\begin{figure}
	\includegraphics[width=\columnwidth]{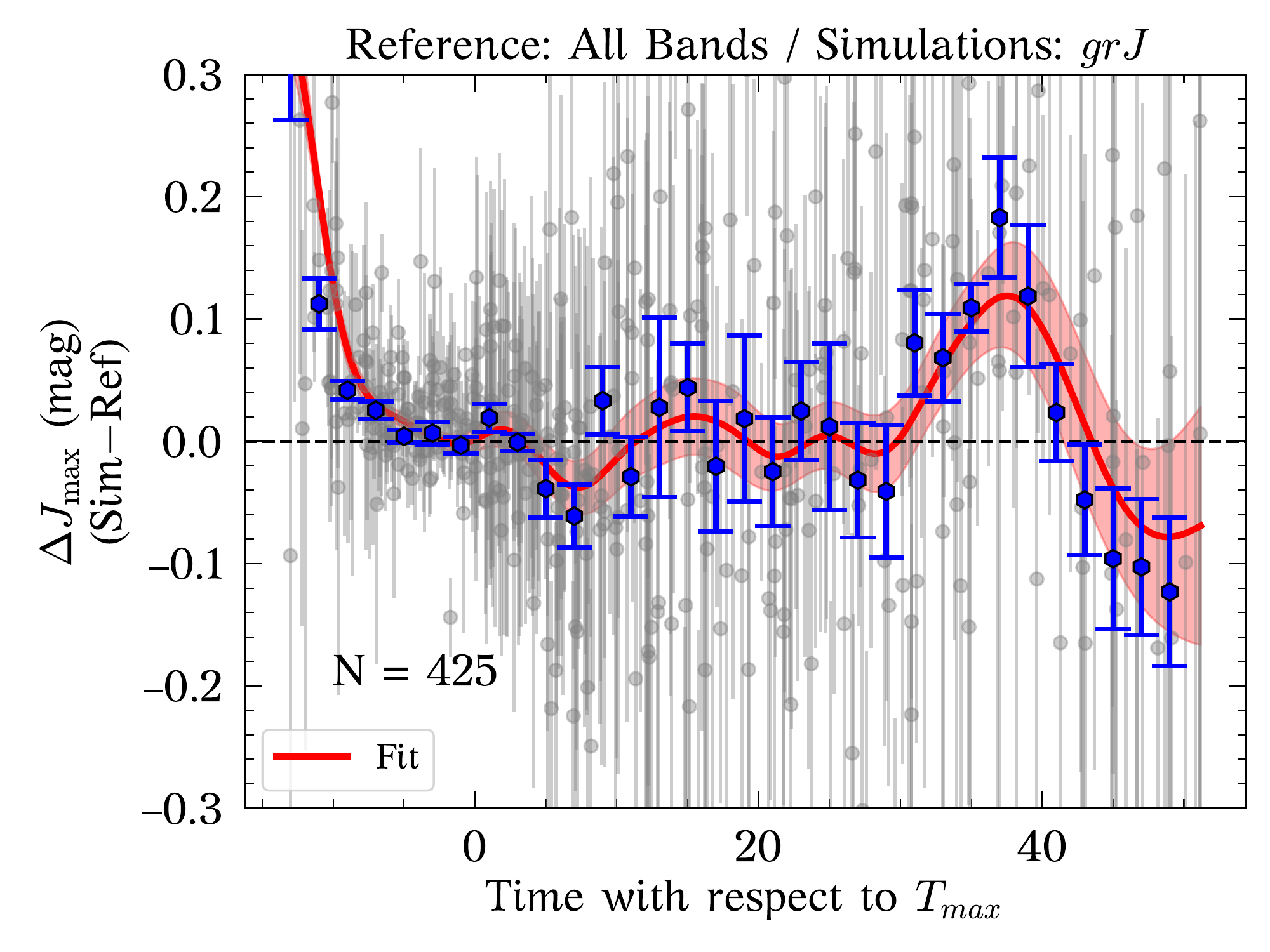}
	\includegraphics[width=\columnwidth]{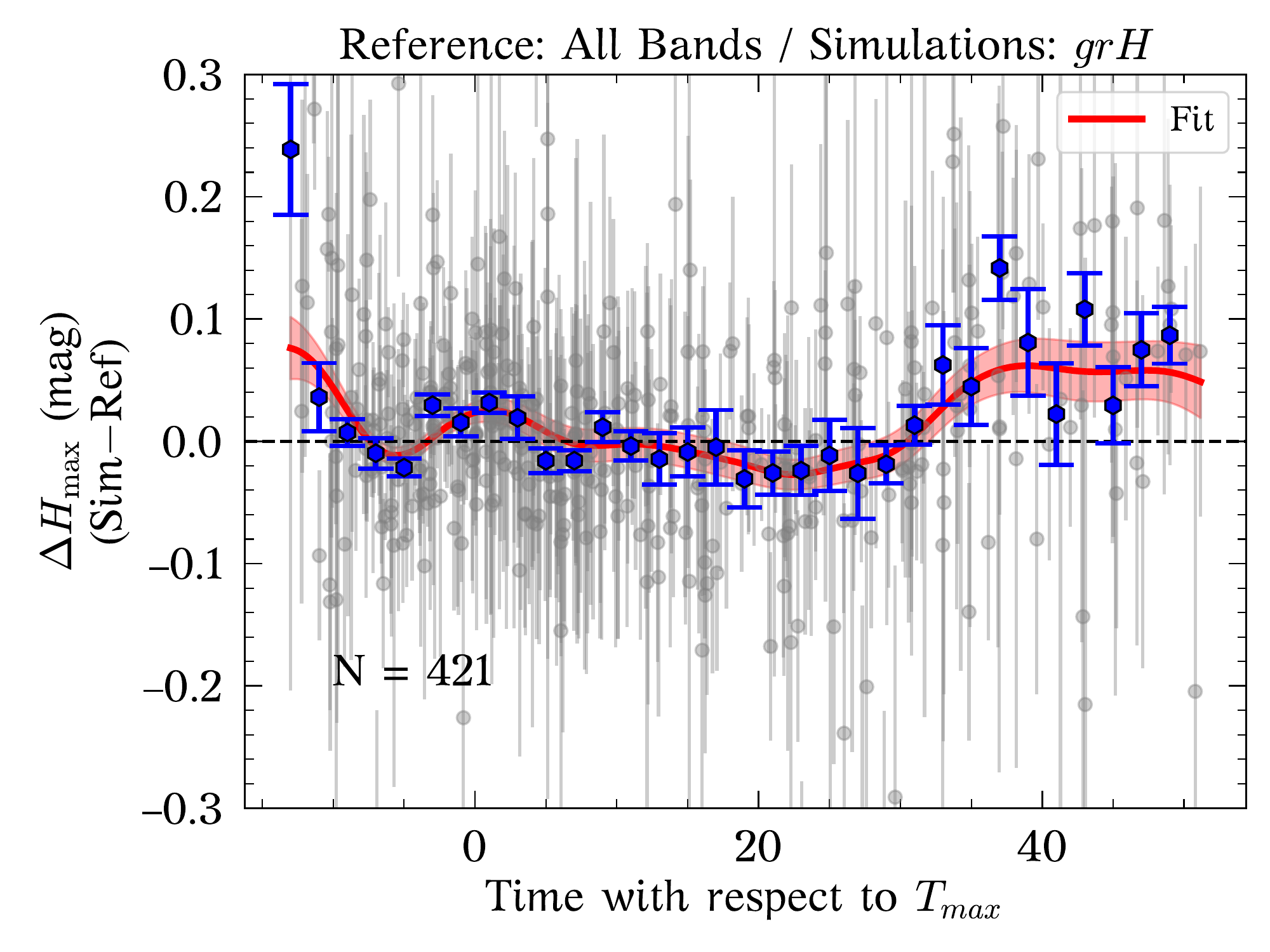}
    \caption{\Jmax (\textit{top} panel) and \Hmax (\textit{bottom} panel) residuals, between simulations with 1 NIR epoch, using \textit{grJ} and \textit{grH}, respectively, and reference values. The weighted mean ($\Delta$) and uncertainty on the weighted mean ($\sigma$) in bins of 2 days are shown in blue. A ``correction snake\rq\rq\ and its uncertainty are calculated by fitting the residuals with GPs (red line and shaded region). N is the total number of simulations.}
    \label{fig:residuals_app}
\end{figure}

In Figure~\ref{fig:residuals_app}, we show show the residuals in \Jmax and \Hmax using all bands for the reference sample, and \textit{grJ} and \textit{grH}, respectively, for the simulations.

\end{document}